\documentclass[a4paper,11pt]{article}
\usepackage[utf8]{inputenc}
\usepackage[T1]{fontenc}
\usepackage[english]{babel}
\usepackage{lmodern,float,csquotes,indentfirst}
\usepackage{amsmath,amssymb,amsfonts,bm,physics}
\usepackage[squaren]{SIunits}
\usepackage{setspace}
\usepackage{graphicx}
\usepackage[textfont={footnotesize,sf},labelfont={color=blue,bf,sf},labelsep=endash]{caption}
\usepackage[position=top,labelfont={color=blue,bf,sf}]{subfig}
\usepackage[pdfpagelabels,colorlinks=true,linkcolor=blue,citecolor=blue,urlcolor=blue]{hyperref}
\usepackage{cite}

\providecommand{\pacs}[1]{\noindent PACS numbers: #1\\}
\providecommand{\keywords}[1]{\noindent Keywords: #1}
\usepackage[affil-sl]{authblk}
\title{\textbf{\Large Transmission in Graphene through Time-oscillating Linear Barrier}}
\author[a]{El Bou\^azzaoui Choubabi}

\author[a,b]{Ahmed Jellal\footnote{a.jellal@ucd.ac.ma}}
\author[a]{Miloud Mekkaoui}
\affil[a]{Laboratory of Theoretical Physics, Faculty of Sciences, Choua\"ib Doukkali University,}
\affil[ ]{PO Box 20, 24000 El Jadida, Morocco}
\affil[b]{Saudi Center for Theoretical Physics, Dhahran, Saudi Arabia}
\date{\small}
\providecommand{\pacs}[1]{\noindent PACS numbers: #1\\}
\providecommand{\keywords}[1]{\noindent Keywords: #1}
\begin{document}
\hypersetup{pageanchor=false}
\begin{titlepage}
	\maketitle
	\thispagestyle{empty}
	\vspace{3cm}

 	\begin{abstract}
 Transmission probabilities of Dirac fermions 
 in graphene under  linear barrier potential
 oscillating in time 
 {are} investigated. Solving Dirac equation
 we end up with the solutions of the energy spectrum depending
 on several modes coming from the oscillations. These will be
 used to obtain a transfer matrix that allows to
 determine transmission amplitudes of all modes.
 Due to numerical difficulties in truncating the resulting coupled channel equations, 
 we limit ourselves to low quantum channels, i.e. $l = 0, \pm1$,
 and study the three corresponding transmission probabilities.

 \end{abstract}
	\vspace{5cm}
	\pacs{72.10.-d; 73.63.-b; 72.80.Rj}
	\keywords{Graphene, time-oscillating linear potential, Dirac equation, transmission modes.
}
\end{titlepage}
\hypersetup{pageanchor=true}
\newpage

\section{ Introduction}

Graphene is a stable planar monolayer of densely crystallized carbon atoms in
a two-dimensional honeycomb lattice \cite{Novoselov1}. Since its realization in $2004$, 
graphene has generated 
an immense interest in  {studying} their mechanical, electronic, optical, thermal 
and chemical properties \cite{Novoselov2,Castro,Beenakker}. The electronic properties are justified by
{a relativistic Hamiltonian resulted from 
Tight-Binding model framework} at low energy 
to the neighborhoods of the six Dirac points at the  Brillouin zone corners in the  reciprocal lattice. 
These are represented by two inequivalent points $K$ and $K'$ 
corresponding to the two atoms $A$ and $B$ constituting the pattern in the direct lattice. 
In addition, the {fermions} in graphene behave as chiral massless particles with
a "light speed" equal to the crystal velocity Fermi $(v_F \approx c / 300)$ with a gapless 
linear dispersion relation near Dirac points. This allows graphene to be a candidate for manufacturing
 the carbon-based nanoelectronic devices.

Quantum transport in periodically driven quantum systems is an
important subject not only of academic value but also for device
and optical applications. In particular, quantum interference
within an oscillating time-periodic electromagnetic field gives
rise to additional sidebands at energies $\epsilon + l\hbar
\omega$ $ (l=0,\pm1,\cdots)$ in the transmission probability
originating from the fact that electrons exchange energy quanta
$\hbar \omega $ carried by photons of the oscillating field,
$\omega $ being the frequency of the oscillating field. The
standard model in this context is that of a time-modulated scalar
potential in a finite region of space. It was studied earlier by
Dayem and Martin \cite{Dayem1} who provided the experimental
evidence of photon assisted tunneling in experiments on
superconducting films under microwave fields. Later on,  Tien and
Gordon \cite{Tien1} provided the first theoretical explanation of
these experimental observations. Further theoretical studies were
performed later by many research groups, in particular Buttiker
investigated the barrier traversal time of particles interacting
with a time-oscillating barrier \cite{Buttiker1}. {Wagner
\cite{Wagner-1} gave a detailed treatment on photon-assisted
tunneling through a strongly driven double barrier tunneling diode. He
also studied the transmission probability of electrons traversing a
quantum well subject to a harmonic driving force \cite{Wagner-2}
where transmission side-bands have been predicted. Grossmann
\cite{Grossmann}, on the other hand, investigated the tunneling
through a double-well perturbed by a monochromatic driving force
which gave rise to unexpected modifications in the tunneling
phenomenon.}
{Very recently, theoretical studies have suggested that an
analog of topological-insulating behavior can be induced
in graphene by a time-dependent electric potential \cite{7,8,9,10,11,12}.
This can be realized by exposing graphene to circularly 
polarized electromagnetic radiation of wavelength much
larger than the physical sample size, such that only the
electric field has significant coupling to the electron 
degrees of freedom \cite{13}}.

{In} 
\cite{Bahlouli} we have 
 solved the 2D Dirac equation describing graphene in the presence of a linear vector potential. The
discretization of the transverse momentum due to the infinite mass boundary condition reduced our 2D
Dirac equation to an effective massive 1D Dirac equation with an effective mass equal to the quantized
transverse momentum. We have used both a numerical Poincaré map approach, based on space discretization of
the original Dirac equation, and a direct analytical method 
{to study
tunneling phenomena through a biased graphene strip. It is} 
showed that 
the numerical results generated by the Poincaré
map are in complete agreement with the analytical results.
{In 
\cite{Mekkaoui},
we have analyzed
the energy spectrum of a graphene sheet subject to a 
magnetic field and a single barrier 
oscillating in time}. {The corresponding transmission
is studied as function of the incident energy and potential parameters. 
In particular, it is showed that}  the time-periodic 
electrostatic potential generates additional sidebands at energies in the transmission
probability originating from the photon absorption or emission within the oscillating barrier. 

Based on our previous work \cite{Bahlouli,Mekkaoui},
we consider a graphene sheet 
subjected to {the oscillating linear barrier potential $V(x,t)$} along the
$x$-direction while the carriers are free in the $y$-direction. 
{More precisely,
the barrier height $V_1$ of $V(x,t)$  
oscillates sinusoidally 
with 
the amplitude $U_{1}$ and frequency $\omega$}. After getting the energy spectrum,
we match all wavefunctions
to end with a transfer matrix that {allow to determine
the transmission probabilities for all Floquet side-bands. Since we have many energy modes 
we focus only
on three channels and describe numerically 
the corresponding transmissions in terms of different physical parameters of our
system.}

The manuscript is organized as follows. In section 2, we
present the theoretical model describing a graphene sheet in the
presence of  {a linear barrier potential oscillating in time}. Subsequently,  we
explicitly determine the eigenvalues and 
eigenspinors for each region composing our system. 
The transmission probabilities for all energy modes will be determined using
the transfer matrix approach in section 3.
%
{In section 4, we numerically present and  
discuss  the transmissions for three channels under suitable
conditions 
of the physical parameters
characterizing our system.
We
conclude our results in the final section}.

\section{ Theoretical formulation}

\begin{figure}[!h]
\centering
\includegraphics[width=9cm, height=7cm]{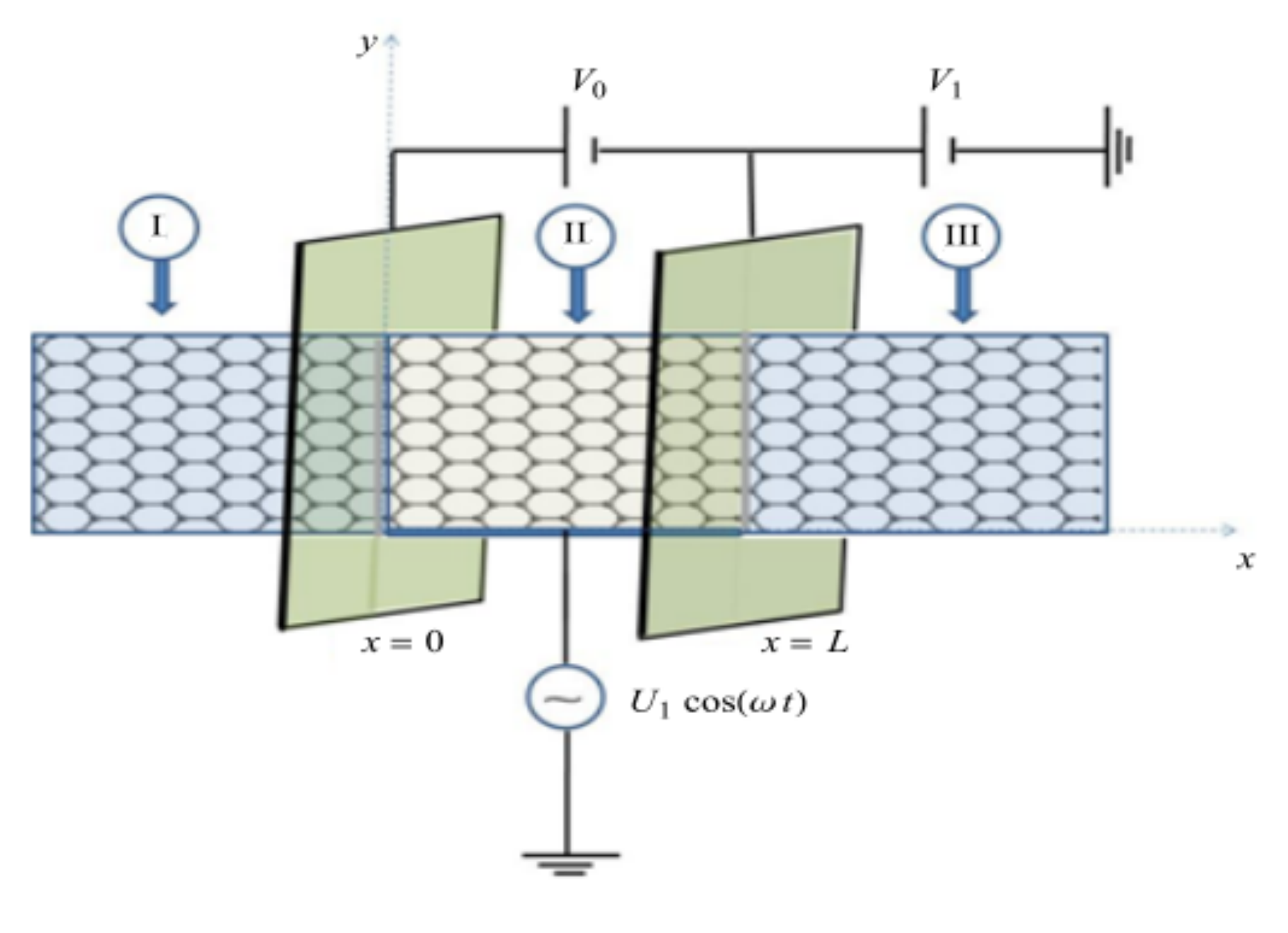}\
 \caption{\sf{
    (Color online)
   Schematic of a graphene sheet whose region II subjected to  
   a time-oscillating linear barrier  potential.}}\label{figure00}
\end{figure}

\begin{figure}[!ht]
\centering
\includegraphics[width=9cm, height=7cm]{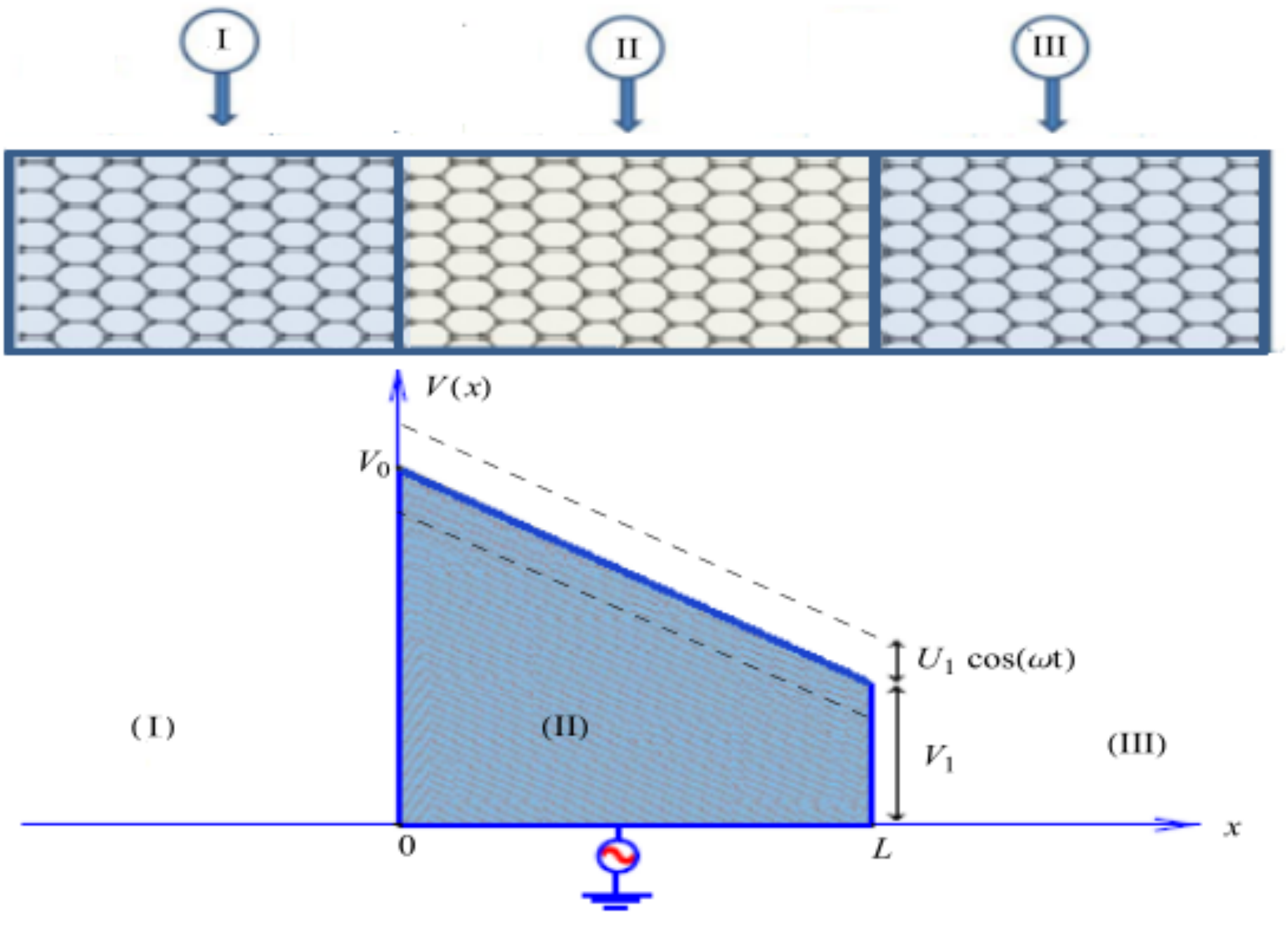}\
 \caption{\sf{
    (Color online)
   Configuration of time-oscillating linear barrier potential on graphene sheet 
   composed of 
   of three regions {(I,II,III)}. }} \label{figure01}
\end{figure}

{We consider a  system composed of three regions 
(Figure \ref{figure00}) where
in  regions  (I, III) there is only pristine graphene and
the intermediate region II 
is subjected 
time-harmonic potential of amplitude $U_1$ 
and driving frequency $\omega$} as well as linear potential generated by 
the two flat armatures sited at  interfaces $ x = 0 $ and $ x = L $. 
The armatures are identical, with the same surface 
spaced by a distance $ L $, and  are perpendicular to the graphene sheet. 
The condenser (two flat armatures) is biased by a potential difference $V_0-V_1$. 
We assume that 
the dimensional requirement   allows the condenser  
to be considered as infinite in order to neglect 
the edge effects. 
In the condenser, the electric field prevailing between the two armatures comes from 
superposition of the electric fields created by each armature, the resultant field remains 
perpendicular to the armatures.
Since {such field} spatially derives of a potential so the  equipotential surfaces are parallel to 
the  armatures and vary gradually linear between the two potentials reported in the armatures. 
We assume that the graphene sheet passes through grooves at the  centers of armatures without 
disturbing the system (Figure \ref{figure00}) in such a way that the  intermediate region of 
the sheet is subjected to a linear inter-armatures potential (Figure \ref{figure01}) given by 
\begin{eqnarray}
V (x,t) &=& - \left(\frac{V_{0}-V_{1}}{L}\right)x + V_0 + U_1 \cos(\omega t)\nonumber \\
&=& -Sx+V_0+ U_1 \cos(\omega t), \qquad  V_{0}>V_{1}.
\end{eqnarray}

\noindent
Note that our system is 
made of 
massless Dirac fermions moving  along the $x$-direction 
and being free in the $y$-direction.
We assume that 
the graphene sheet is  characterized by  very large length scale ($x$-direction) and width $W$ 
($y$-direction). 
In input region I, at the interface $x=0$, the incident fermions of energy $E_0$  
and angle $\phi_{0}$ with respect to the $x$-direction are reflected with 
energies 
$E_0+ m\hbar \omega$ $(m=0, \pm 1, \pm 2, \cdots)$ and incident angles $\pi-\phi_{m}$. 
In output region III,  at the interface $x=L$, after transmission the fermions have the 
same energies with transmission angle  $\phi_{m}$. 

The Hamiltonian 
{system} 
can be written as
\begin{equation}\label{ham1}
H=-i\hbar v_{F} \sigma \cdot \nabla+V(x,t){\mathbb I}_{2}
\end{equation}
where $\upsilon_{F}$ is the Fermi velocity, $ \sigma =(\sigma_{x},
\sigma_{y})$ are the usual Pauli matrices and ${\mathbb I}_{2}$
is the $2 \times 2$ unit matrix (hereafter we use $v_{F}=\hbar=1$).  
{The} system has finite width $W$ with infinite mass
boundary conditions on the wavefunction at  boundaries $y = 0$
and $y = W$ along the $y$-direction \cite{Tworzydlo, Berry}. 
These result in a quantization of the transverse
momentum 
\begin{equation}
k_{y}=\frac{\pi}{W}\left(n+\frac{1}{2}\right), \qquad n=0,1,2 \cdots.
\end{equation}
On other hand, 
{the periodic nature of the linear potential 
necessitates that the solutions of the Dirac fermions  are 
Floquet states.
Therefore, we write 
the eigenspinors as
$\psi(x , y, t) 
=\psi(x, y)\chi(t)$}
and then from the eigenvalue equation
$H\psi_j(x,y,t)=  i \partial_t \psi_j(x,y,t)$
we obtain
\begin{equation}
\left[E_0+U_{1}\cos\left(\omega t\right)\right]\psi_j(x,y,t)= i
\partial_t \psi_j(x,y,t)
\end{equation}
which can be integrated to end up with the solution
\begin{eqnarray}
 \psi_j(x,y,t) 
= \psi_j(x,y,0)e ^{-i E_0 t} e ^{-i U_{1}\sin\left(\omega t\right)
/ \omega}.
\end{eqnarray}
By introducing the expansion 
in terms  of the $m$-th order Bessel function of the first
kind  $J_{m}$, 
we can write the eigenspinors as
\begin{eqnarray}\label{ch0}
 \psi_j(x,y,t)= \psi_j(x,y,0)e ^{-i E_0 t}\sum_{m=-\infty}^{+\infty}J_m (\alpha)e ^{i m\omega t}
\end{eqnarray}
{corresponding to the eigenvalues  including the Floquet side-bands} 
\begin{equation}
E=E_0+m\omega
\end{equation}
where we have set  $\alpha=\frac{U_{1}}{\omega}$. 
Taking into account of the energy conservation, the eigenspinors that
describe the fermions in the $j$-th region can be expressed as a
linear combination of those  at energies $E+l\omega$
$(l=0,\pm 1,\pm 2, \cdots)$. Thus we have
\begin{equation}\label{ch2}
\psi_{ j}(x,y,t)=e^{ik_{y}y}\sum^{m,l=+\infty}_{m,l=-\infty}
\psi_{j}^{l}(x,y) J_{m-l}\left(\alpha\right)e^{-i(E+m\omega)t}
\end{equation}
 {and the spinor
$\psi_{j}^{l}(x,y)$ will be determined  by 
considering  each region $j$}. 
Indeed, in region I, solving 
the time-dependent Dirac equation 
we can easily get 
the solution 
at energy $E$ for the {incident} 
fermions
\begin{eqnarray}
&& \psi_{inc}(x,y,t)=\left(
\begin{array}{c}
1 \\
 \alpha_{0}\end{array}\right)e^{ik_{0}x}e^{ik_{y}y}e^{-iEt}
\\
&& \alpha_{0}=s_{0}\frac{k_{0} +ik_{y}}{\sqrt{k_{0}^{2}
+k_{y}^{2}}}=s_{0} e^{\textbf{\emph{i}}\phi_{0}}
\end{eqnarray}
where $s_{0}=\mbox{sgn}(E)$, $\phi_{0}$ is the angle that
the incident fermions make with the $x$-direction,  $k_{0}$
and $k_{y}$ are the $x$ and $y$-components of the fermion wave
vector, respectively. {Because of the oscillation in time of $V(x,t)$, then the reflected and transmitted waves have
components at all energies $E+ l\omega$. 
Consequently the eigenspinors $\psi_{r}(x,y,t)$ for reflected
fermions are}
\begin{eqnarray}
&& \psi_{r}(x,y,t)=\sum^{+\infty}_{m,l=-\infty} r_{l}\left(
\begin{array}{c}
1 \\
 -\frac{1}{\alpha_{l}}\end{array}\right)e^{-ik_{l} x +ik_{y}
 y}  J_{m-l} \left(\alpha\right) e^{-i(E+m\omega)t}\\
 &&
 \alpha_{l}=s_{l}\frac{k_{l} +ik_{y}}{\sqrt{k^{2}_{l}
+k_{y}^{2}}}=s_{l} e^{\textbf{\emph{i}}\phi_{l}}
\end{eqnarray}
and correspond to the eigenvalues
\begin{equation}\label{energy1}
E+l\omega=s_{l}\sqrt{k^{2}_{l}+k^{2}_{y}} 
\end{equation}
where $r_{l}$ is {the amplitude of reflection} and
$J_{m-l} \left(\alpha\right)=\delta_{m,l}$ because  the modulation amplitude is $V_{j} = 0$
in
this case. 
$\phi_{l}=\tan^{-1}(k_{y}/k_{l})$, sign $s_{l}=\mbox{sgn}(E+l
\omega)$ refers to the conduction and valence
bands of region I. The number $k_l$ can be obtained from
\eqref{energy1}
\begin{equation}
k_{l}=s_{l}\sqrt{\left(E+l\omega\right)^{2}-k^{2}_{y}}.
\end{equation}
{Finally combing all to obtain the eigenspinors in
region  I $(x<0)$}
\begin{equation}
\psi_{\text{I}}(x,y,t)=e^{ik_{y}y}\sum^{+\infty}_{m,l=-\infty}\left[\delta_{l,0}\left(
\begin{array}{c}
1 \\
 \alpha_{l}\end{array}\right)e^{ik_{l}x}+r_{l}\left(
\begin{array}{c}
1 \\
 -\frac{1}{\alpha_{l}}\end{array}\right)e^{-ik_{l} x
 }\right]\delta_{m,l}e^{-i(E+m\omega)t}.
\end{equation}

For region II $(0<x<L)$ 
a linear combination of eigenspinors at energies
$E+l\omega$ 
has to be taken and therefore
we have
\begin{equation}
 \psi_{\text{II}}(x,y,t)=\sum^{+\infty}_{l=-\infty}\psi_{l}(x,y)
 \times\sum^{+\infty}_{m=-\infty} J_{m} \left(\alpha\right) e^{-im\omega
t-i(E+l\omega)t}
\end{equation}
{which}
can be written in terms of the parabolic
cylinder function \cite{Bahlouli} 
such that  the first component
is given by
\begin{equation}\label{hi1}
 \phi^{+}_{\text{II}}=c_{1}
 D_{\nu-1}\left(Q_{l}\right)+c_{2}
 D_{-\nu}\left(-Q^{*}_{l}\right)
\end{equation}
{corresponding to the energies
\begin{equation}
\epsilon_{l}=E+l\omega-V_{0}
\end{equation}
where}
$\nu=\frac{ik_{y}^{2}}{2S}$,
$Q_{l}(x)=\sqrt{\frac{2}{S}}e^{i\pi/4}(Sx+\epsilon_{l})$, $c_{1}$
and $c_{2}$ are constants. The second component takes the form
\begin{eqnarray}\label{hi2}
\phi^{-}_{\text{II}}
=-\frac{c_{2}}{k_{y}}\left[ 2(\epsilon_{l}+Sx)
 D_{-\nu}\left(-Q^{*}_{l}\right)
+
 \sqrt{2S}e^{i\pi/4}D_{-\nu+1}\left(-Q^{*}_{l}\right)\right]
 -\frac{c_{1}}{k_{y}}\sqrt{2S}e^{-i\pi/4}
 D_{\nu-1}\left(Q_{l}\right).
\end{eqnarray}
The components of the spinor solution of the Dirac equation \eqref{ham1}
in region II can be {derived from \eqref{hi1} and \eqref{hi2} as}
\begin{eqnarray}
\psi_{ II}(x,y,t) 
&=& \left(
\begin{array}{c}
 \phi^{+}_{\text{II}}+i\phi^{-}_{\text{II}} \\
 \phi^{+}_{\text{II}}-i\phi^{-}_{\text{II}}\\
\end{array}%
\right)
\\
&=& e^{ik_{y}y}\sum^{+\infty}_{l=-\infty}c_{1,l}\left(%
\begin{array}{c}
 \eta^{+}_{l}(x) \\
  \eta^{-}_{l}(x) \\
\end{array}%
\right)+c_{2,l}\left(%
\begin{array}{c}
 \xi^{+}_{l}(x) \\
 \xi^{-}_{l}(x)\\
\end{array}%
\right) 
\sum^{+\infty}_{m=-\infty}J_{m-l}\left(\alpha\right)
e^{-i(E+m\omega)t}\nonumber
\end{eqnarray}
{where we have defined  the functions $ \eta^{\pm}_{l}(x)$ and  $\xi^{\pm}_{l}(x)$}
\begin{eqnarray}
&& \eta^{\pm}_{l}(x)=
 D_{\nu-1}\left(Q_{l}\right)\mp
 \frac{\sqrt{2S}}{k_{y}}e^{i\pi/4}D_{\nu}\left(Q_{l}\right)\\
&& \xi^{\pm}_{l}(x)=
 \pm\frac{1}{k_{y}}\sqrt{2S}e^{-i\pi/4}D_{-\nu+1}\left(-Q_{l}^{*}\right)
  \pm
 \frac{1}{k_{y}}(-2i\epsilon_{l}\pm
 k_{y}-2iS x)D_{-\nu}\left(-Q_{l}^{*}\right).
\end{eqnarray}

For region {III} $(x>L)$, the eigenspinors
$\psi_{t}(x,y,t)$ for transmitted fermions read as
\begin{equation}
\psi_{t}(x,y,t)=\sum^{+\infty}_{m,l=-\infty}t_{l}\left(
\begin{array}{c}
1 \\
 \alpha_{l}\end{array}\right)e^{ik_{l} x +ik_{y}
 y}\ J_{m-l} \left(\alpha\right)e^{-i(E+m\omega)t}
\end{equation}
where $t_{l}$ is {the  amplitude transmission, which can be mapped 
in terms of the null vector  $\{b_{l}\}$ as}
\begin{equation}
\psi_{\text{III}}(x,y,t)=e^{ik_{y}y}\sum^{+\infty}_{m,l=-\infty}\left[t_{l}\left(
\begin{array}{c}
1 \\
 \alpha_{l}\end{array}\right)e^{ik_{l} x
 }+b_{l}\left(
\begin{array}{c}
1 \\
 -\frac{1}{\alpha_{l}}\end{array}\right)e^{-ik_{l} x}\right]\delta_{m,l}e^{-i(E+m\omega)t}.
\end{equation}
{In the forthcoming analysis, we will see how to use
the above solutions of the energy spectrum to deal with 
different issues. 
In fact, we will employ them 
to explicitly
determine the transmission probabilities for all energy modes}.

\section{Transmission through oscillating linear barrier}

Using the orthogonality of $\{e^{im\omega
t}\}$ and the continuity of eigenspinors at boundaries ($x=0$, $x=L$), 
to end up with a set of equations for 
$r_m$, $t_m$, $b_m$, $c_{1,m}$, $c_{2,m}$ 
\begin{eqnarray}
&& \delta_{m,0}+r_{m}=\sum^{+\infty}_{l=-\infty}
\left(c_{1,l}\eta_{l}^{+}(0) +c_{2,l}\xi_{l}^{+}(0)\right)
J_{m-l}\left(\alpha\right)\\
&& \delta_{m,0}\alpha_{m}-r_{m}\frac{1}{\alpha_{m}}=
\sum^{+\infty}_{l=-\infty}
\left(c_{1,l}\eta_{l}^{-}(0)+c_{2,l}\xi_{l}^{-}(0)\right)
J_{m-l}\left(\alpha\right)\\
&& t_{m}e^{ik_{m}L}+ b_{m}e^{-ik_{m}L}=\sum^{+\infty}_{l=-\infty}
\left(
 c_{1,l}\eta_{l}^{+}(L) +c_{2,l}\xi_{l}^{+}(L)\right) J_{m-l}\left(\alpha\right)\\
&&t_{m}\alpha_{m}e^{ik_{m}L}-b_{m}\frac{1}{\alpha_{m}}e^{-ik_{m}L}
= \sum^{+\infty}_{l=-\infty} \left(c_{1,l}\eta_{l}^{-}(L)
-c_{2,l}\xi_{l}^{-}(L)\right)J_{m-l}\left(\alpha\right).
\end{eqnarray}
Since Dirac fermions pass through a region subjected to
time-oscillating linear potential, transitions from the central band to
sidebands (channels) at energies $E+ m\omega$ $(m = 0, \pm 1,
\pm 2, \cdots)$ occur as fermions ex-change energy quanta with the
oscillating field. This procedure is most conveniently expressed
in the transfer matrix formalism, such as 
\begin{eqnarray}\label{Mat1}
\left(%
\begin{array}{c}
  \Xi_{0} \\
  \Xi_{0}^{'} \\
\end{array}%
\right)=\left(%
\begin{array}{cc}
 { \mathbb M11} &{\mathbb M12} \\
 {\mathbb M21} &{ \mathbb M22} \\
\end{array}%
\right)\left(%
\begin{array}{c}
  \Xi_{2} \\
  \Xi_{2}^{'}\\
\end{array}%
\right)={\mathbb M}\left(%
\begin{array}{c}
  \Xi_{2} \\
 \Xi_{2}^{'} \\
\end{array}%
\right)
\end{eqnarray}
where the total transfer matrix ${\mathbb M}={\mathbb
M(0,1)}\cdot{\mathbb M(1,2)}$ and transfer
matrices ${\mathbb M(j,j+1)}$, that couple the wave function in the $j$-th region to the
wave function in the $(j + 1)$-th region, are
\begin{eqnarray}
&& {\mathbb M(0,1)}=\left(%
\begin{array}{cc}
  {\mathbb I}& {\mathbb I} \\
{\mathbb N_{0}^{+}} &{\mathbb N_{0}^{-}} \\
\end{array}%
\right)^{-1}
\left(%
\begin{array}{cc}
  {\mathbb C_{0}^{+}} & {\mathbb G_{0}^{+}} \\
 {\mathbb C_{0}^{-}} & {\mathbb G_{0}^{-}} \\
\end{array}%
\right)\\
&& {\mathbb M(1,2)}=\left(%
\begin{array}{cc}
  {\mathbb C_{L}^{+}} & {\mathbb G_{L}^{+}} \\
  {\mathbb C_{L}^{-}} & {\mathbb G_{L}^{-}} \\
\end{array}%
\right)^{-1}
\left(%
\begin{array}{cc}
  {\mathbb I}& {\mathbb I} \\
{\mathbb N_{0}^{+}} &{\mathbb N_{0}^{-}} \\
\end{array}%
\right)\left(%
\begin{array}{cc}
  {\mathbb K^{+}}&{\mathbb O}  \\
{\mathbb O} &{\mathbb K^{-}} \\
\end{array}%
\right)
\end{eqnarray}
by setting the quantities 
\begin{eqnarray}
&& \left({\mathbb
N_{0}^{\pm}}\right)_{m,l}=\pm\left(\alpha_{m}\right)^{\pm
1}\delta_{m,l}, \qquad 
\left({\mathbb
C_{\tau}^{\pm}}\right)_{m,l}=\eta_{l}^{\pm}(\tau)J_{m-l}\left(\alpha\right)\\
&& \left({\mathbb
G_{\tau}^{\pm}}\right)_{m,l}=\xi_{l}^{\pm}(\tau)J_{m-l}\left(\alpha\right), \qquad 
\left({\mathbb  K^{\pm}}\right)_{m,l}=\pm e^{\pm
iLk_{m}}\delta_{m,l}
\end{eqnarray}
where 
the null matrix is denoted by ${\mathbb O}$ and  ${\mathbb I}$ is
the unit matrix. We are considering fermions propagating from left to
right with energy $E$, then, $\tau=(0,L)$,
$\Xi_{1}=\{\delta_{0,l}\}$ and $\Xi_{2}^{'}=\{b_{m}\}$ is the null
vector, whereas $\Xi_{2}=\{t_{l}\}$ and $\Xi_{1}^{'}=\{r_{l}\}$
are the vectors of transmitted  and reflected waves
respectively. From the above considerations, one can easily obtain
the relation
\begin{equation}
\Xi_{2}=\left({ \mathbb M11}\right)^{-1}\ \Xi_{1}.
\end{equation}
The minimum number $N$ of sidebands that need to be considered is
determined by the strength of the oscillation,
$N>\frac{U_{1}}{\omega}$. Then the infinite series for $T$ can
be truncated to consider a finite number of terms starting from
$-N$ up to $N$. Furthermore, analytical results are obtained if we
consider small values of $\alpha=\frac{U_{1}}{\omega}$ and
include only the first two sidebands at energies $E\pm
\omega$ along with the central band at energy $E$. This gives the result
\begin{equation}
t_{-N+k}=\mathbb M^{'}\left[k+1, N+1\right]
\end{equation}
where $k=0, 1, 2,\cdots,N$ and ${ \mathbb M^{'}}$ denotes 
the inverse matrix $\left({ \mathbb M11}\right)^{-1}$. Then for
$\alpha=0$, it remains only the transmission $t_{0}=\mathbb M^{'}\left[1,1\right]$ for central
bands, which    gives
exactly the result obtained in our previous work \cite{Bahlouli}
\begin{equation}
t_{0}= 
\frac{e^{-ik_{0}L}\left[1+z_{0}^{2}\right]\left[\xi^{+}(L)\eta^{-}(L)-\xi^{-}(L)\eta^{+}(L)\right]}
{\left[\xi^{+}(0)+z_0\xi^{-}(0)\right]\left[\eta^{-}(L)-z_0\eta^{+}(L)\right]-
\left[\eta^{+}(0)+z_0\eta^{-}(0)\right]\left[\xi^{-}(L)-z_0\xi^{+}(L)\right]}.
\end{equation}

On the other hand, we can use
 the reflected
$J_{\sf {ref}}$ and transmitted $J_{\sf {tra}}$ currents to {explicitly determine}
 the   transmission $T_{l}$ and reflection $R_{l}$ probabilities
 corresponding to our system. {These are defined as}
\begin{equation}\label{ttrr}
  T_{l}=\frac{ |J_{{\sf {tra}},l}|}{| J_{{\sf {inc}},0}|},\qquad 
  R_{l}=\frac{|J_{{\sf {ref}},l}|}{ |J_{\sf {inc,0}}|}
\end{equation}
{$T_{l}$ describes} the
scattering of fermions with incident energy $E$ in the
region I into the sideband with energy $E+l\omega$ in
region III. Thus, the rank of the transfer matrix ${\mathbb
M}$ increases with the amplitude of the time-oscillating
potential. {To evaluate \eqref{ttrr}, we use  the Hamiltonian system to show
that
the current density takes the form} 
\begin{equation}
J= v_{F}\psi^{\dagger}\sigma _{x}\psi
\end{equation}
{and now using the solutions of the energy spectrum for each region to end up with
the incident, reflected and transmitted currents} 
\begin{eqnarray}
&&J_{{\sf {inc}}, 0}= v_{F}(\alpha_{0}+(\alpha_{0})^{\ast})\\
&&  J_{{\sf {ref}}, l}= v_{F}r_{l}^{\ast}r_{l}(\alpha_{l}+(\alpha_{l})^{\ast})\\ 
 && J_{{\sf {tra}}, l}= v_{F}t_{l}^{\ast}t_{l}(\alpha_{l}+(\alpha_{l})^{\ast}).
\end{eqnarray}
{Injecting them into \eqref{ttrr} to obtain}
\begin{equation}
  T_{l}= \frac{k_{l}}{k_{0}}
  | t_{l}|^{2},\qquad  R_{l}= \frac{k_{l}}{k_{0}}
  | r_{l}|^{2}
\end{equation}
Due to numerical difficulties, we  truncate 
\eqref{Mat1} retaining only  terms corresponding to
the central and first sidebands, namely $l = -1, 0, 1$, {which are} 
\begin{equation}
  t_{-1}=\mathbb M^{'}[1,2], \qquad t_{0}=\mathbb M^{'}[2,2], \qquad t_{1}=\mathbb M^{'}[3,2].
\end{equation}
To explore the above results and go deeply in order to underline our system behavior, we will
introduce the numerical analysis. For this, we will focus only on few channels and choose different
configurations of the physical parameters.

\section{Numerical results}

\begin{figure}[!ht]
\centering
\subfloat[]{
    \includegraphics[scale=0.42]{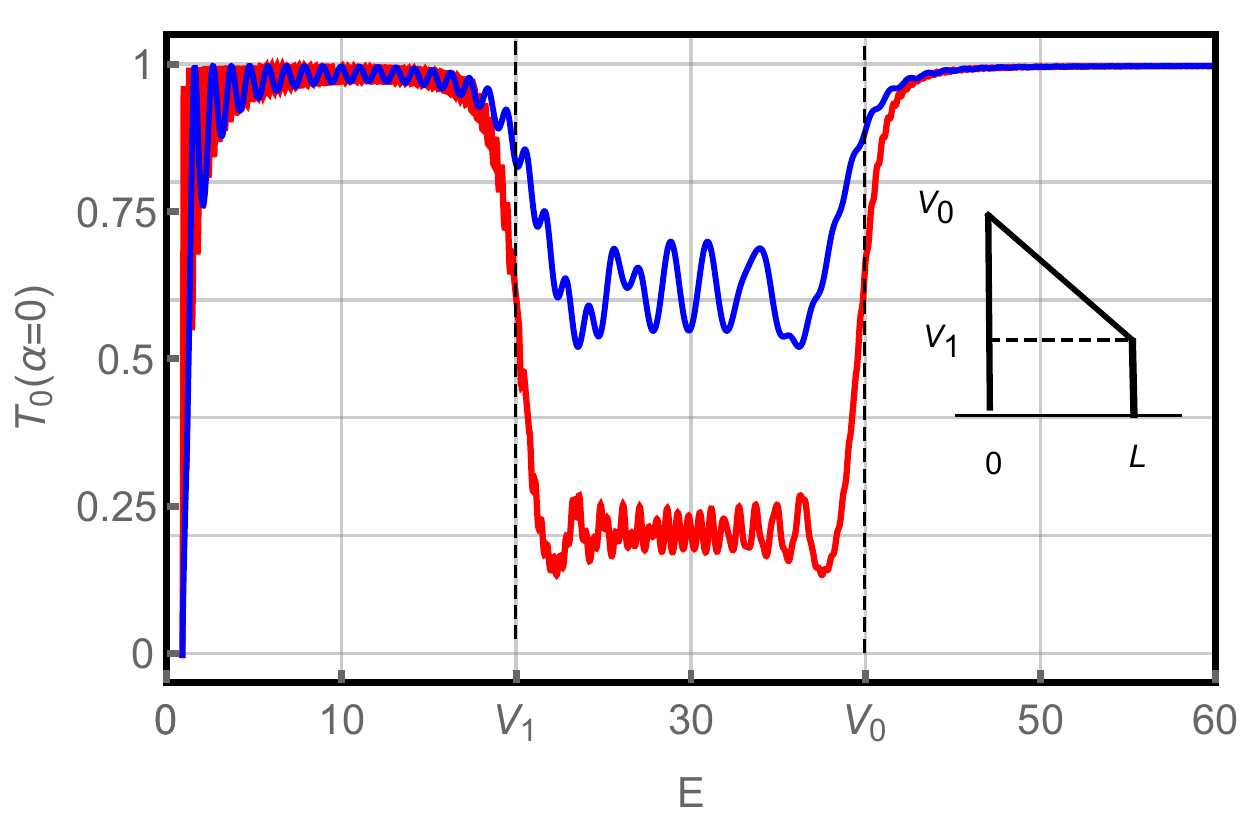}
    \label{FigCh01:100}
}
\subfloat[]{
    \includegraphics[scale=0.42]{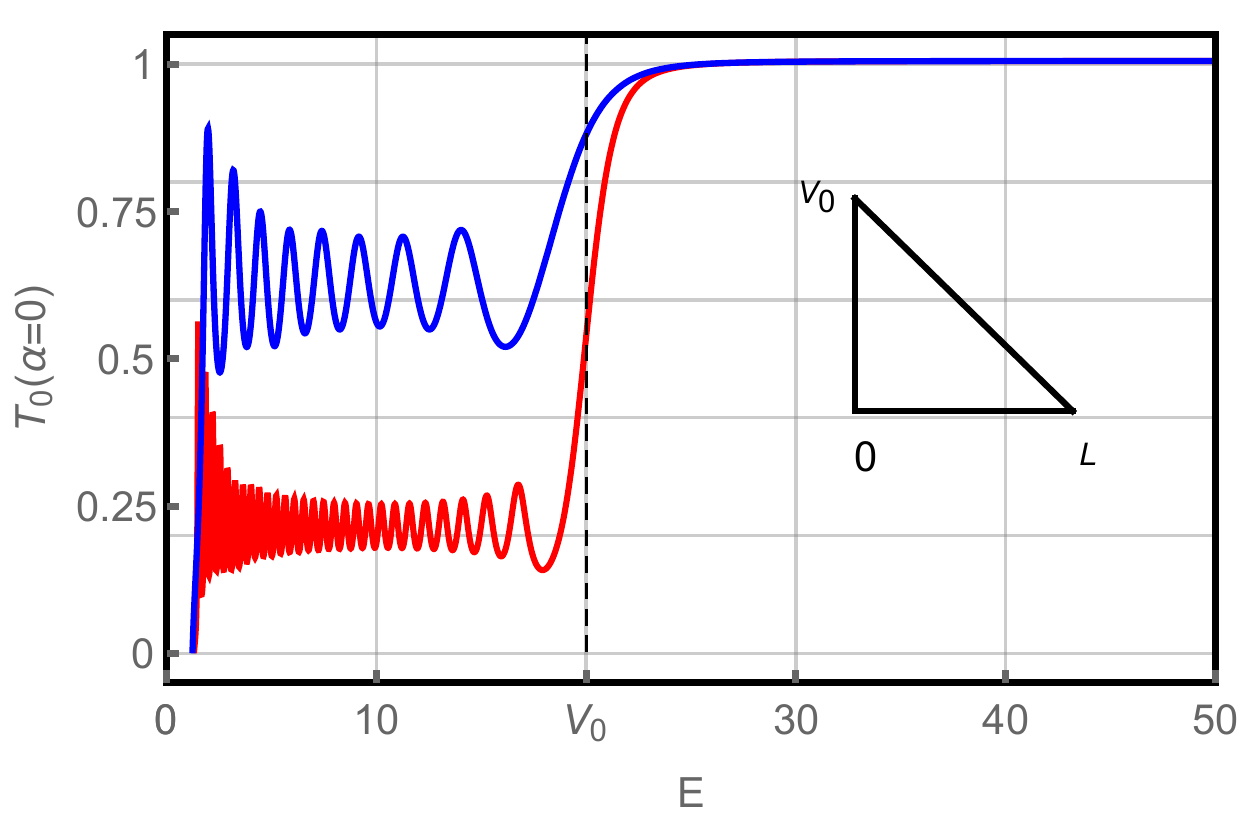}
    \label{FigCh01:101}
}
\subfloat[]{
    \includegraphics[scale=0.42]{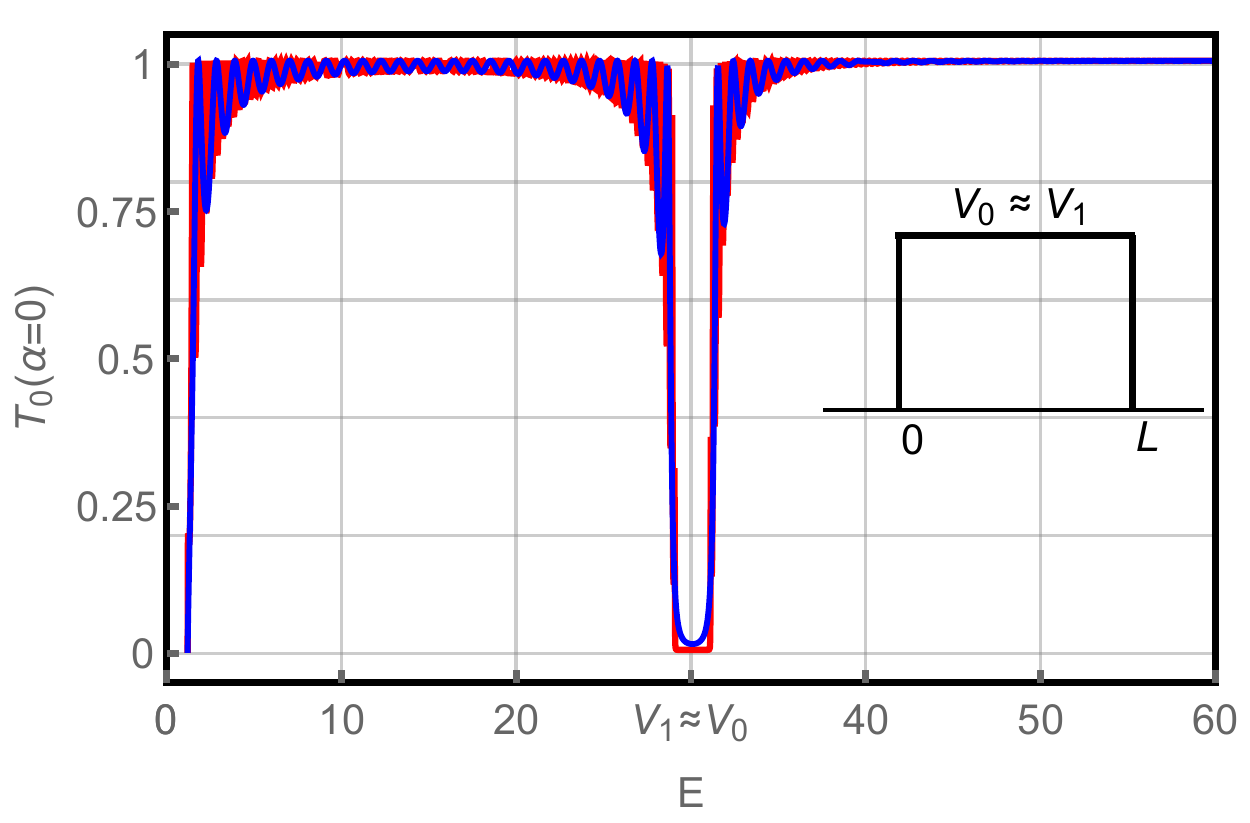}
    \label{FigCh01:102}
}\\
\caption{(Color online) {Transmission probability versus} 
incident energy
 $E$, with  $k_{y}=1$, $\alpha=0$, $L=3$ (blue color), $L=10$ (red color)  
 for \protect\subref{FigCh01:100}: $V_{0}=40$ and $V_{1}=20$, 
 \protect\subref{FigCh01:101}: $V_{0}=20$ and $V_{1}=0$ and 
 \protect\subref{FigCh01:102}: $V_{0}\rightarrow V_{1}=30$. }
\label{FigCh01}
\end{figure}

In Figure \ref{FigCh01}, we present the transmission probability versus
incident energy $E$. Figure \ref{FigCh01}\subref{FigCh01:100} shows the general behavior 
of the transmission probability of a linear barrier $ (\alpha=0)$ limited
by two values of potential ($ V_0=40$, $ V_1=20$) for $ L = 3$  and $ L =10 $, respectively. 
We observe that the transmission is forbidden below the energy $ E = ky=1 $, which behaves 
like an effective mass $ ky = m ^ * $ \cite{choubabi1, choubabi2}. For energies 
$ ky \leq E \leq V _1 - 2k y $ 
the transmission shows oscillations and gives rise to the Klein paradox, i.e.  $T_0(\alpha=0) = 1$ (Klein zone). 
In the range of energy $ V _1 - 2k_y \leq E \leq V _0 + 2k_ y$, the transmission decreases 
from $ E=V _1 - 2k_y $ until a minimum value where it oscillates and after increases by passing 
from $ E=V _0 $ to reach a maximal value at $ E =V _0 + 2k_ y $. However  for energy $ E >V _0 + 2k_ y $ 
our system shows a classical behavior. 
To give comparison and show the relevance of our finding, 
we plots two cases according to choice of the barrier heights $(V_{0}, V_{1})$. Indeed,  
Figure \ref{FigCh01}\subref{FigCh01:101} illustrates a particular case of  a
linear barrier ($V_{0}=20$, $V_{1}=0$) studied in our previous work \cite{Bahlouli} where  
transmission corresponding 
to the Klein zone  is omitted and 
 transmission oscillates around a minimum, then it behaves in the same way as shown in 
 \ref{FigCh01}\subref{FigCh01:100}. 
Figure \ref{FigCh01}\subref{FigCh01:102} presents the case of a simple  square barrier $V_{0}\rightarrow V_{1}=30$
where
the  Klein zone  is conserved and  transmission corresponding to energies 
$ V _1 - 2k_y \leq E \leq V _0 + 2k_ y$ is replaced by another in range $ V _1 - k_y \leq E \leq V _1+ k_y$ 
without oscillations  \cite{choubabi3}. Finally, we observe that  Figures \ref{FigCh01} highlight the  effect of 
the  barrier width on transmission, as long as $L$ increases the minimal transmission of the intermediate 
zones decreases  
and the number of oscillations increases.

\begin{figure}[!ht]
\centering
\subfloat[]{
    \includegraphics[scale=0.28]{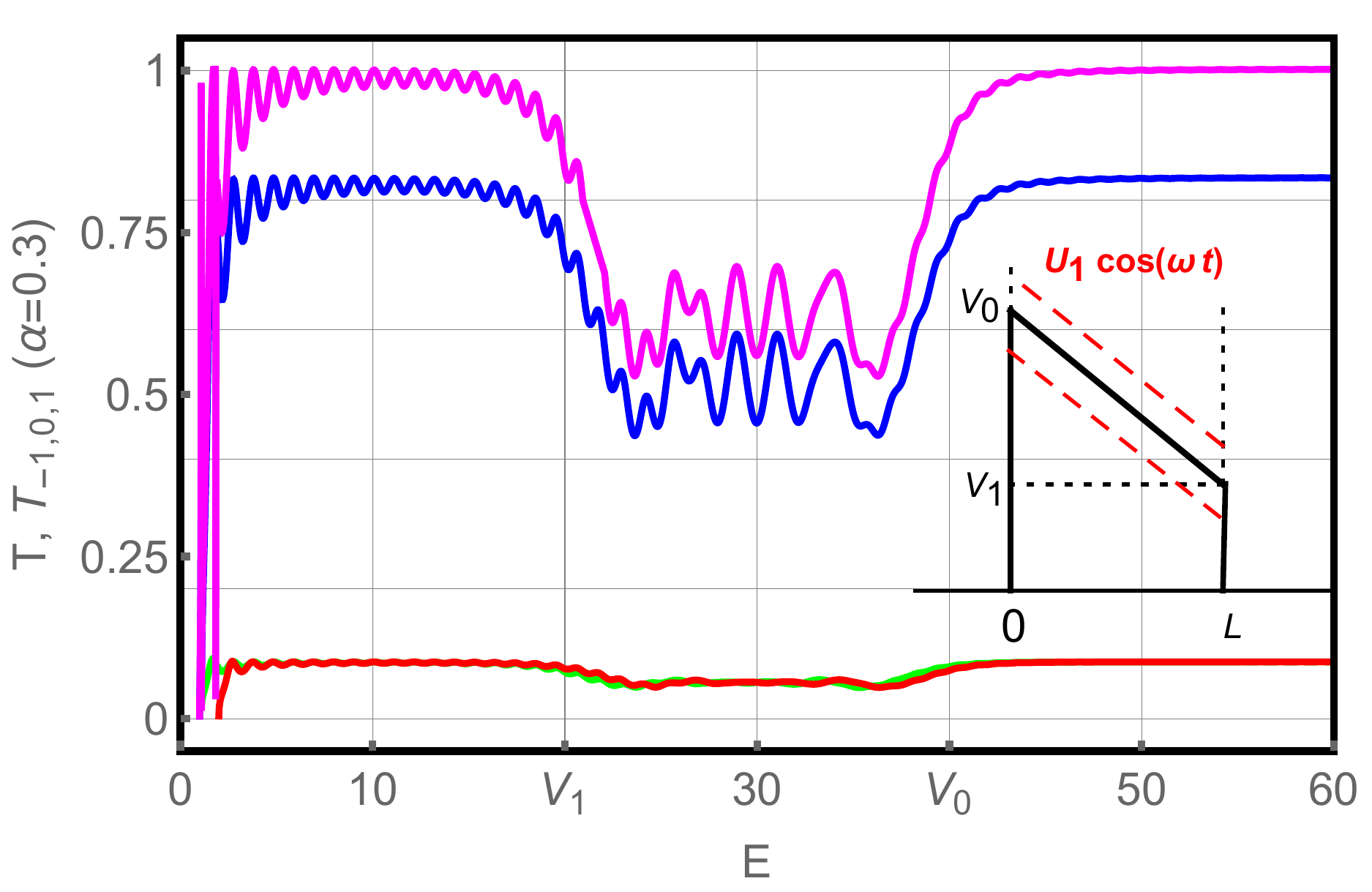}
    \label{FigCh02:200}
}
\subfloat[]{
    \includegraphics[scale=0.28]{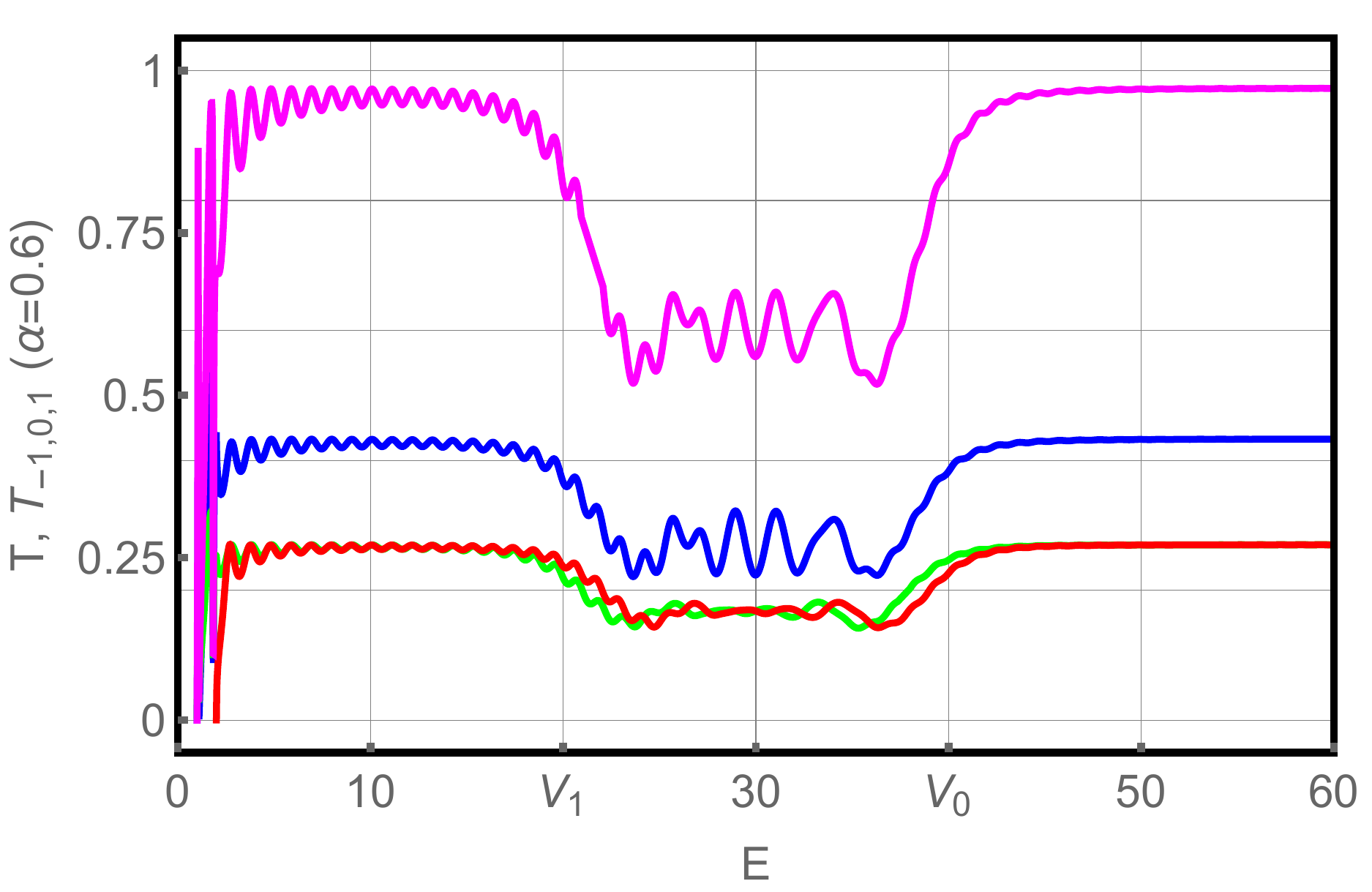}
    \label{FigCh02:201}
}
\subfloat[]{
    \includegraphics[scale=0.28]{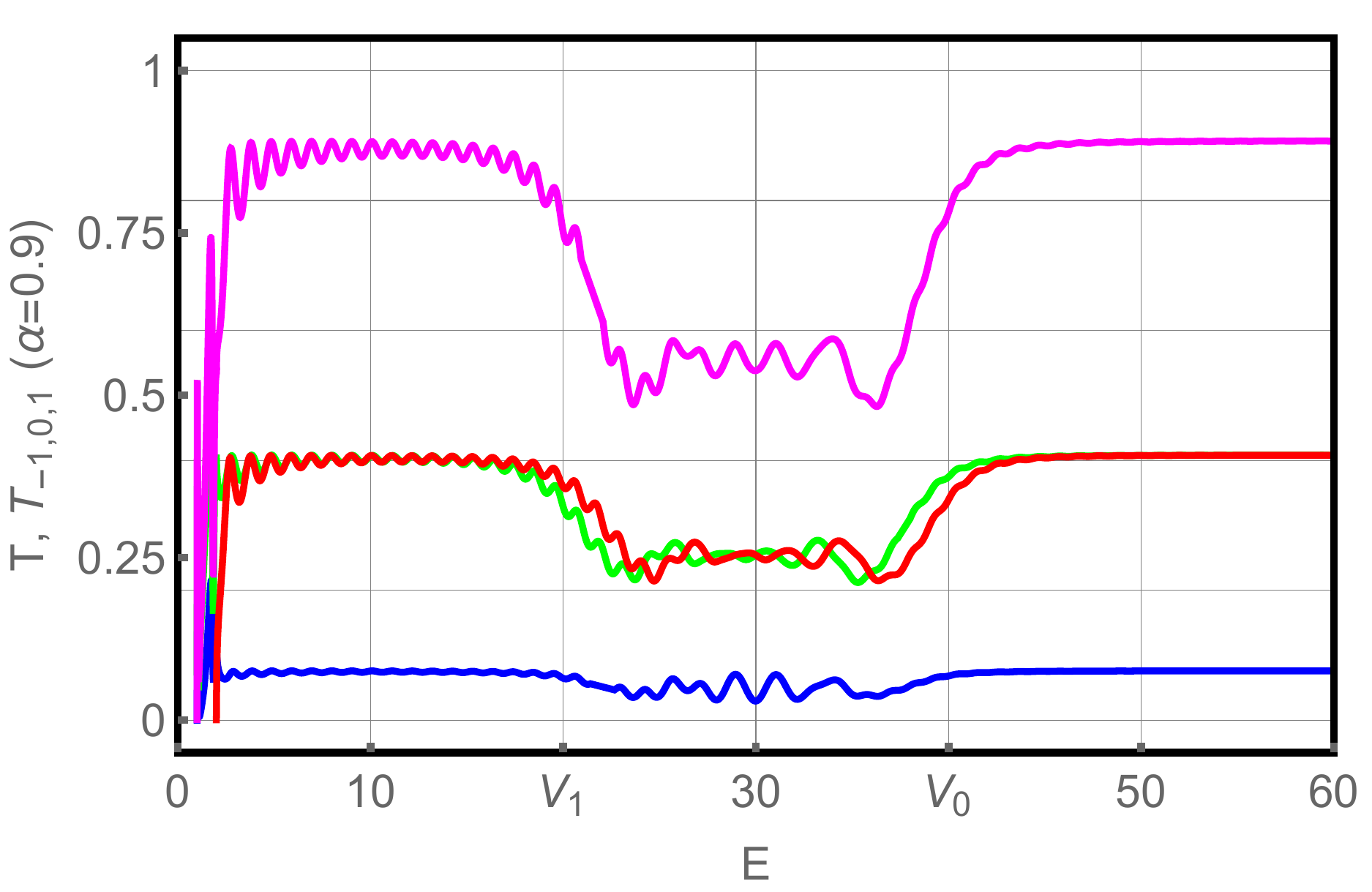}
    \label{FigCh02:202}
}
\caption{(Color online) {Transmission probabilities versus} incident energy
 $E$, with  $k_{y}=1$, $L=3$, $V_1=20$, $V_0=40$ and $\omega=1$  for \protect\subref{FigCh02:200}: $\alpha=0.3$, 
 \protect\subref{FigCh02:201}: $\alpha=0.6$ and \protect\subref{FigCh02:202}: $\alpha=0.9$. $T_{0}$ (blue color), 
 $T_{l=-1}$ (red color), $T_{l=1}$ (green color),   $T$ (magenta color).}
\label{FigCh02}
\end{figure}

In Figures \ref{FigCh02} we present 
the three channel transmissions together with total one (summation over three channels)
versus incident energy $E$ for $\alpha=0.3$, $\alpha=0.6$ and $\alpha=0.9$. These concern
the central transmission band
$T_{0}$ (blue color), two first sidebands $T_{l=-1}$ (red color), $T_{l=1}$ (green color), 
and total $T$
(magenta color) 
  \begin{equation}
  T= \sum_{l=-1, 0, 1}\frac{k_{l}}{k_{0}}
  | t_{l}|^{2}.
  \end{equation}
 Since the time-oscillating linear barrier vibrates sinusoidally and longitudinally along  the propagating
 $x$-direction of Dirac fermions, with amplitude $U_{1}$ and frequency $\omega$, we notice
 that this causes  a change  in
 the effective mass from $k_y$ to 
 $k_{y}\pm\omega$. 
 For low values of $\alpha$ the central
transmission band $T_0$ is dominant compared to those of lateral \ref{FigCh02}\subref{FigCh02:200}. As long as 
$\alpha$ increases the central transmission band $(l=0)$ decreases while the two
first sidebands $(l=\pm 1)$ increase. In Figure \ref{FigCh02}\subref{FigCh02:201} we have co-dominance of 
the three bands but in Figure \ref{FigCh02}\subref{FigCh02:202}  
the two first sidebands $(l=\pm 1)$ dominate.
We observe that
the total transmission  decreases if $\alpha$ increases. Since $\alpha=\frac{U_1}{\omega}$ is barrier height
dependent,
the decreasing of $T$ might be caused by the missing modes. 

\begin{figure}[!ht]
\centering
\subfloat[]{
    \includegraphics[scale=0.28]{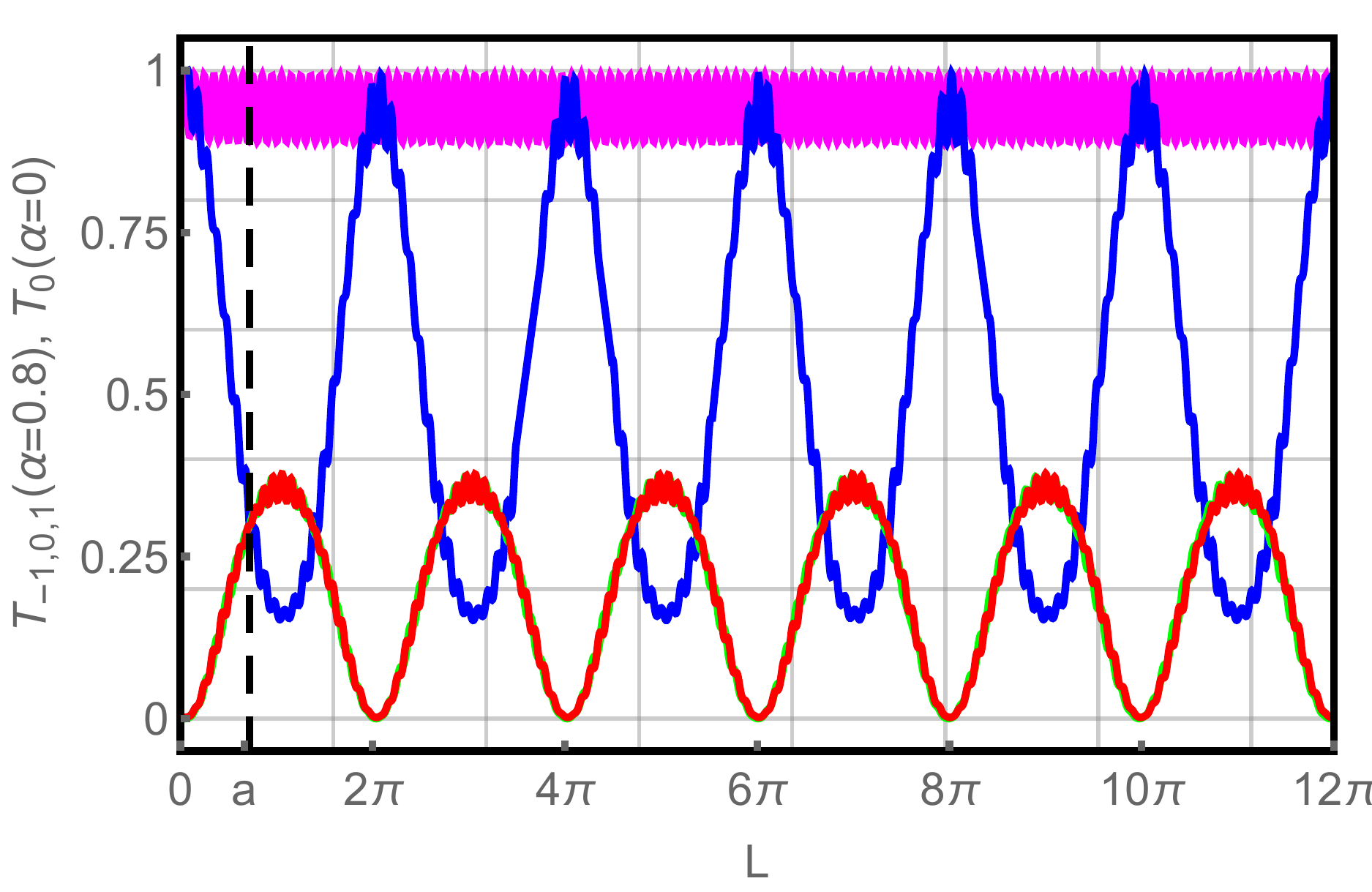}
    \label{FigCh03:301}
}
\subfloat[]{
    \includegraphics[scale=0.28]{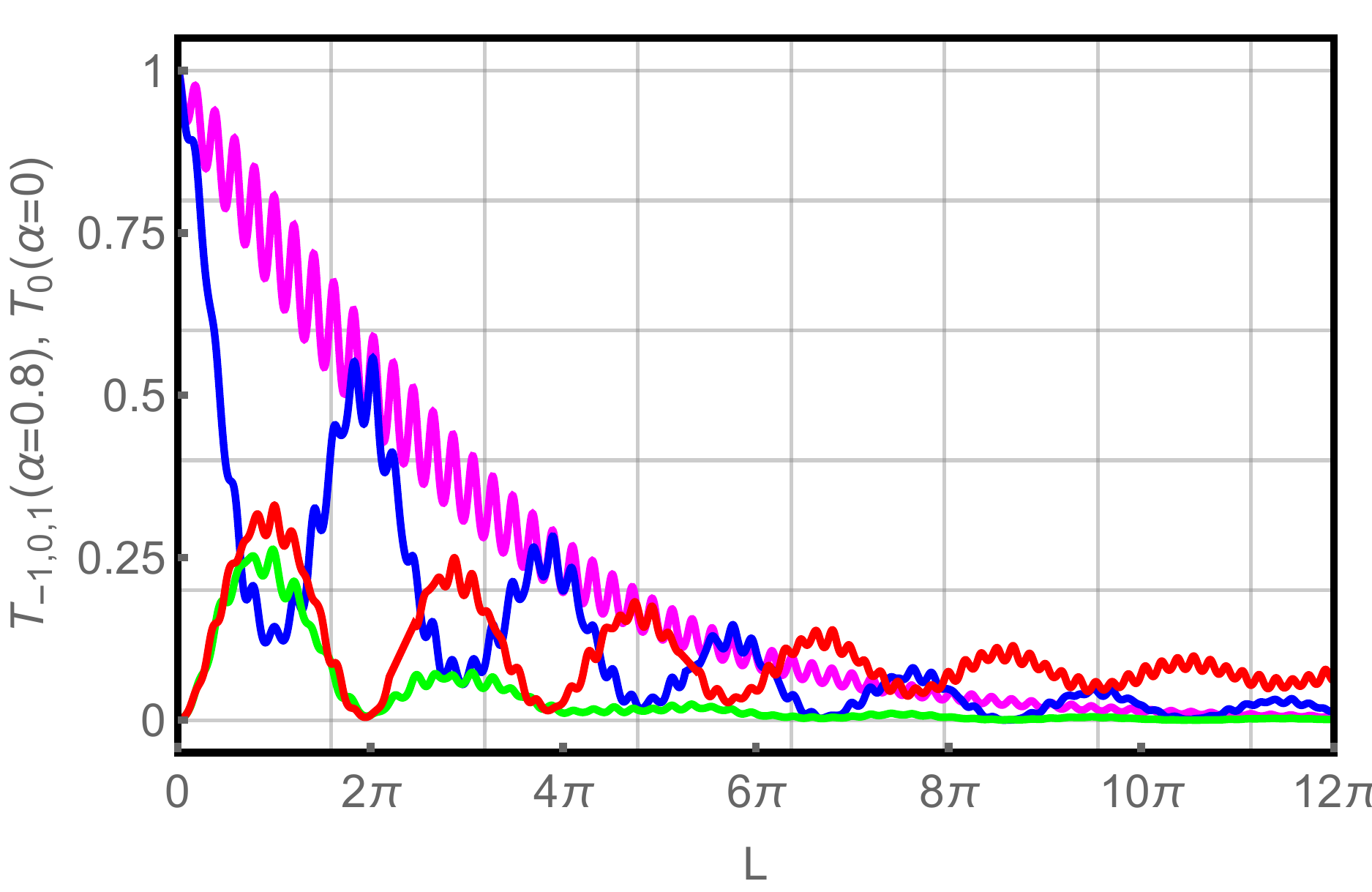}
    \label{FigCh03:302}
}
\subfloat[]{
    \includegraphics[scale=0.28]{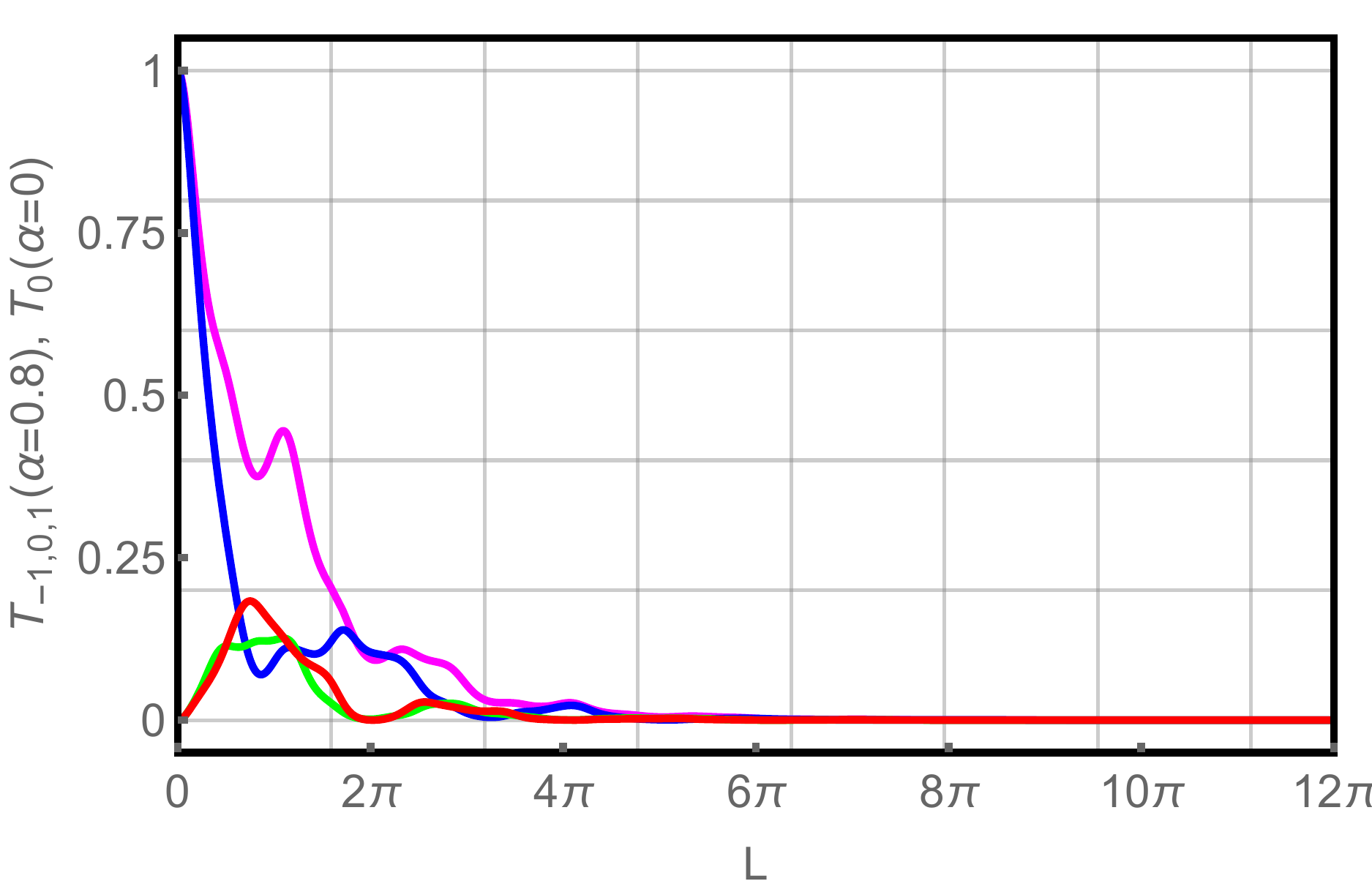}
    \label{FigCh03:303}
}\\
\subfloat[]{
    \includegraphics[scale=0.28]{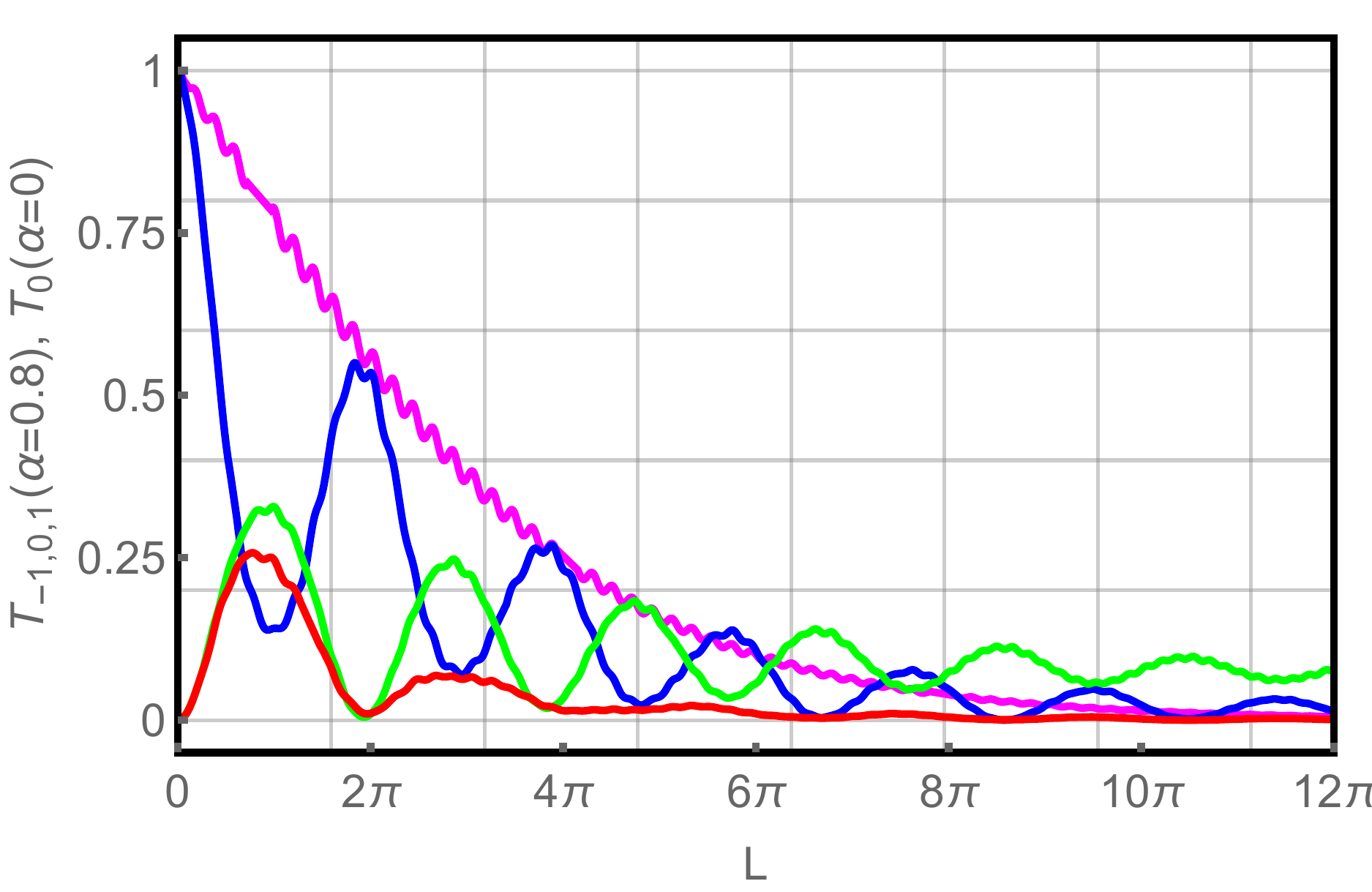}
    \label{FigCh03:304}
}
\subfloat[]{
    \includegraphics[scale=0.28]{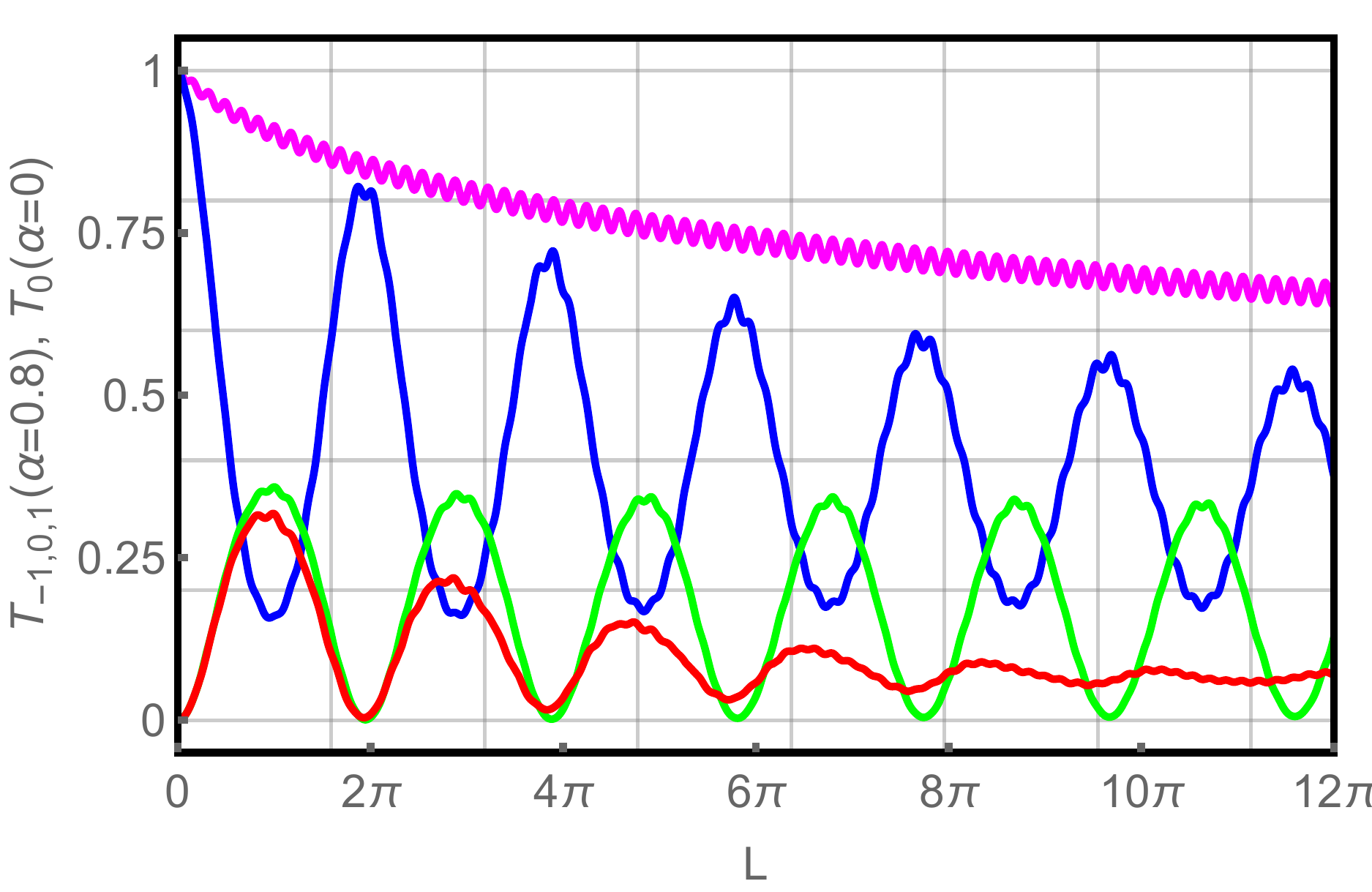}
    \label{FigCh03:305}
}
\subfloat[]{
    \includegraphics[scale=0.28]{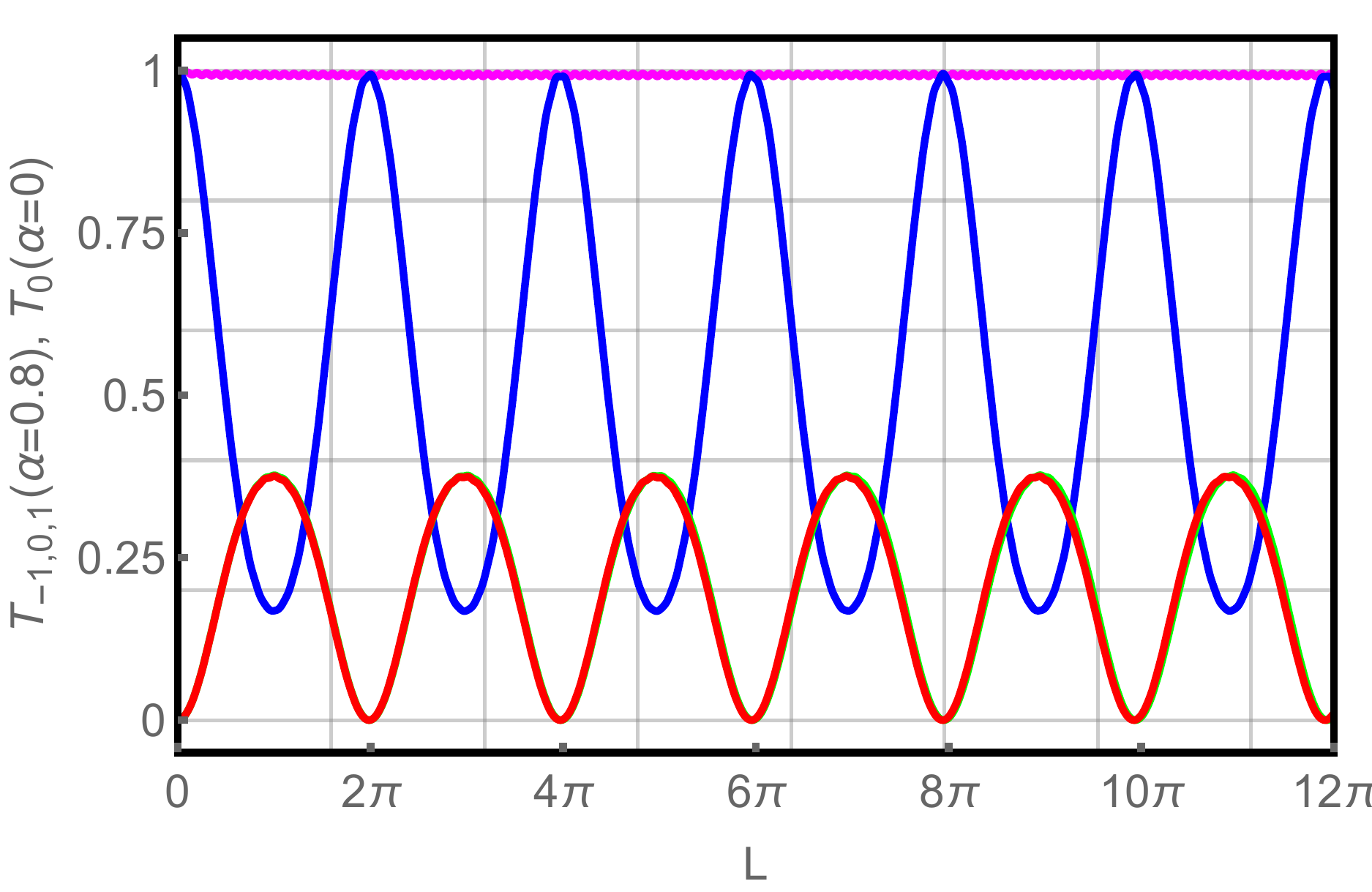}
    \label{FigCh03:306}
}
\caption{(Color online) 
{Transmission probabilities versus barrier width $L$}
with  $k_{y}=1$, $V_1=10$, $V_0=20$, $\alpha=0,0.8$, $T_{0}(\alpha=0)$ (magenta color), 
$T_{0}(\alpha=0.5)$ (blew color), $T_{-1}(\alpha=0.5)$ (red color), and $T_{1}(\alpha=0.5)$ (green color), 
for \protect\subref{FigCh03:301}: $ E=5$, \protect\subref{FigCh03:302}: $ E=10$, \protect\subref{FigCh03:303}: 
$ E=16$, \protect\subref{FigCh03:304}: $ E=20$, \protect\subref{FigCh03:305}: $ E=21$, \protect\subref{FigCh03:306}: 
$ E=25$.}
\label{FigCh33}
\end{figure}

Figure \ref{FigCh33} 
shows transmission
probabilities 
{$T_{-1,0,1}(\alpha=0.8)$ and $T_{0}(\alpha=0))$ as function of 
barrier width $L$ under suitable  conditions. We observe that $T_{-1,0,1}(\alpha=0.8)$} 
have a periodicity of $L=2 \pi$ except $T_{0}(\alpha=0)$. {
In Figure \ref{FigCh33}\subref{FigCh03:301} and for  $0<E<V_1$, transmissions 
maintain sinusoidal oscillations where 
$T_{-1,1}(\alpha=0.8)$ are almost equal throughout the periodicity of $L$. Moreover, we notice that for 
$L\in [2k\pi,2k\pi+a[\cup ]2(k+1)\pi-a,2(k+1)\pi]$, $k\in \mathbb{N}$ and 
$a\in [0,\pi]$, 
$T_{0}(\alpha=0.8)$ is dominating but for $L\in ]2k\pi+a,2(k+1)\pi-a[$ the role changes to 
 $T_{-1,1}(\alpha=0.8)$ and  for $(L=2k\pi+a$ or $L=2(k+1)\pi-a)$} 
$T_{-1,0,1}(\alpha=0.8)$ are co-dominant. Throughout the variation of $L$, $T_{0}(\alpha=0)$ 
alternately takes values close to unity.
Figure \ref{FigCh33}\subref{FigCh03:302} presents 
$T_{-1,0,1}(\alpha=0.8)$ and $T_{0}(\alpha=0)$ for $E=V_1=10$ where the transmissions are waning 
in oscillation as long as $L$  increases. For $L\in [2k\pi,2k\pi+a[\cup ]2(k+1)\pi-a,2(k+1)\pi]$ 
we have the dominance of $T_{0}(\alpha=0.8)$ until $L=8\pi$. Beyond this value, 
$T_{-1}(\alpha=0.8)$ dominates $T_{0}(\alpha=0.8)$ and the two transmissions are maintained 
with the increase of $L$, but $T_{1}(\alpha=0.8)$ vanishes at $L=4\pi$
{and 
$T_{0}(\alpha=0)$} decays sinusoidally and vanishes at $L=10\pi$. In Figure \ref{FigCh33}\subref{FigCh03:303} 
{and for 
$V_1<E<V_0$, we see that all transmission probabilities decay and vanish before $L=5\pi$}. The transmission probabilities 
in Figure \ref{FigCh33}\subref{FigCh03:304} with $E=V_0=20$  are similar to those of 
Figure \ref{FigCh33}\subref{FigCh03:302} with $E=V_1=10$, except that the roles of $T_{-1}(\alpha=0.8)$ 
and $T_{1}(\alpha=0.8)$ are inverted. 
{In Figure \ref{FigCh33}\subref{FigCh03:305}
for
 $E=21>V_0$ 
all transmissions start 
to behave in similar way as in Figure \ref{FigCh33}\subref{FigCh03:301}.}
{In Figure \ref{FigCh33}\subref{FigCh03:306} for $E=25>V_0$, 
$T_{-1,0,1}(\alpha=0.8)$ and $T_{0}(\alpha=0)$   
 are similar to those seen in
 Figure \ref{FigCh33}\subref{FigCh03:301},
the difference 
is that those 
of Figure \ref{FigCh33}\subref{FigCh03:306} are smooth and thin compared to 
those in the Figure \ref{FigCh33}\subref{FigCh03:301}. 
}

\begin{figure}[!ht]
\centering
\subfloat[]{
    \includegraphics[scale=0.28]{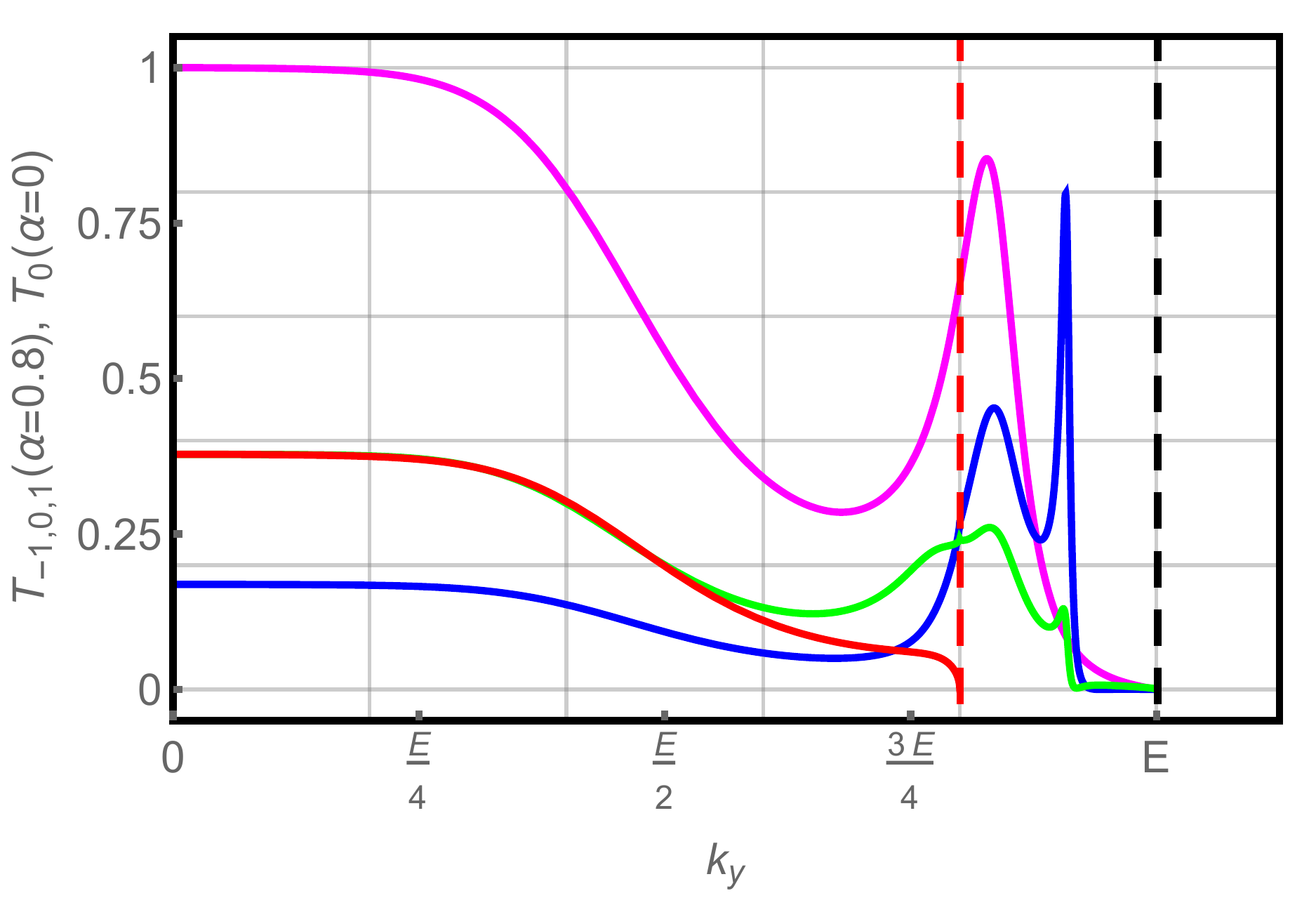}
    \label{FigCh04:401}
}
\subfloat[]{
    \includegraphics[scale=0.28]{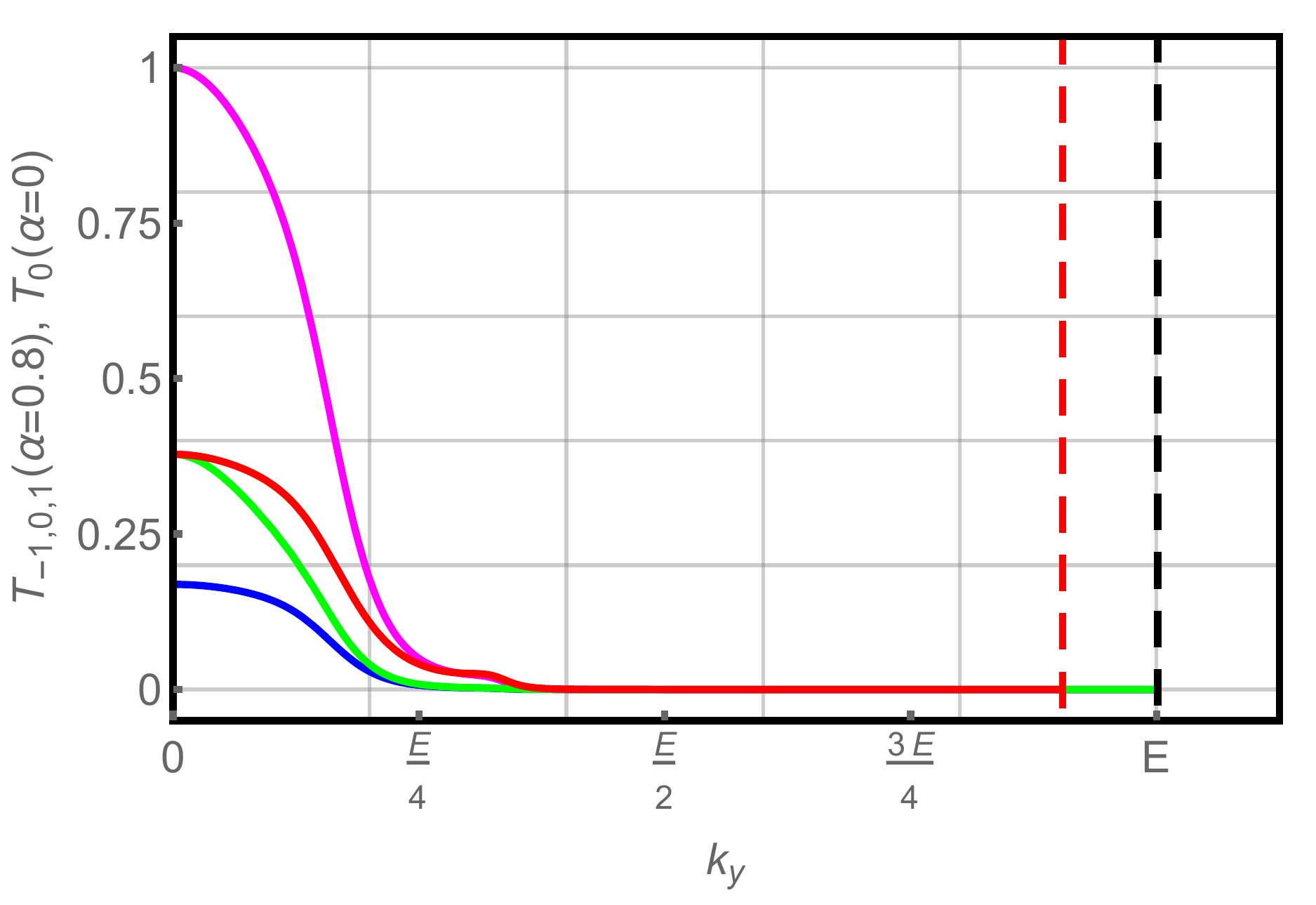}
    \label{FigCh04:402}
}
\subfloat[]{
    \includegraphics[scale=0.28]{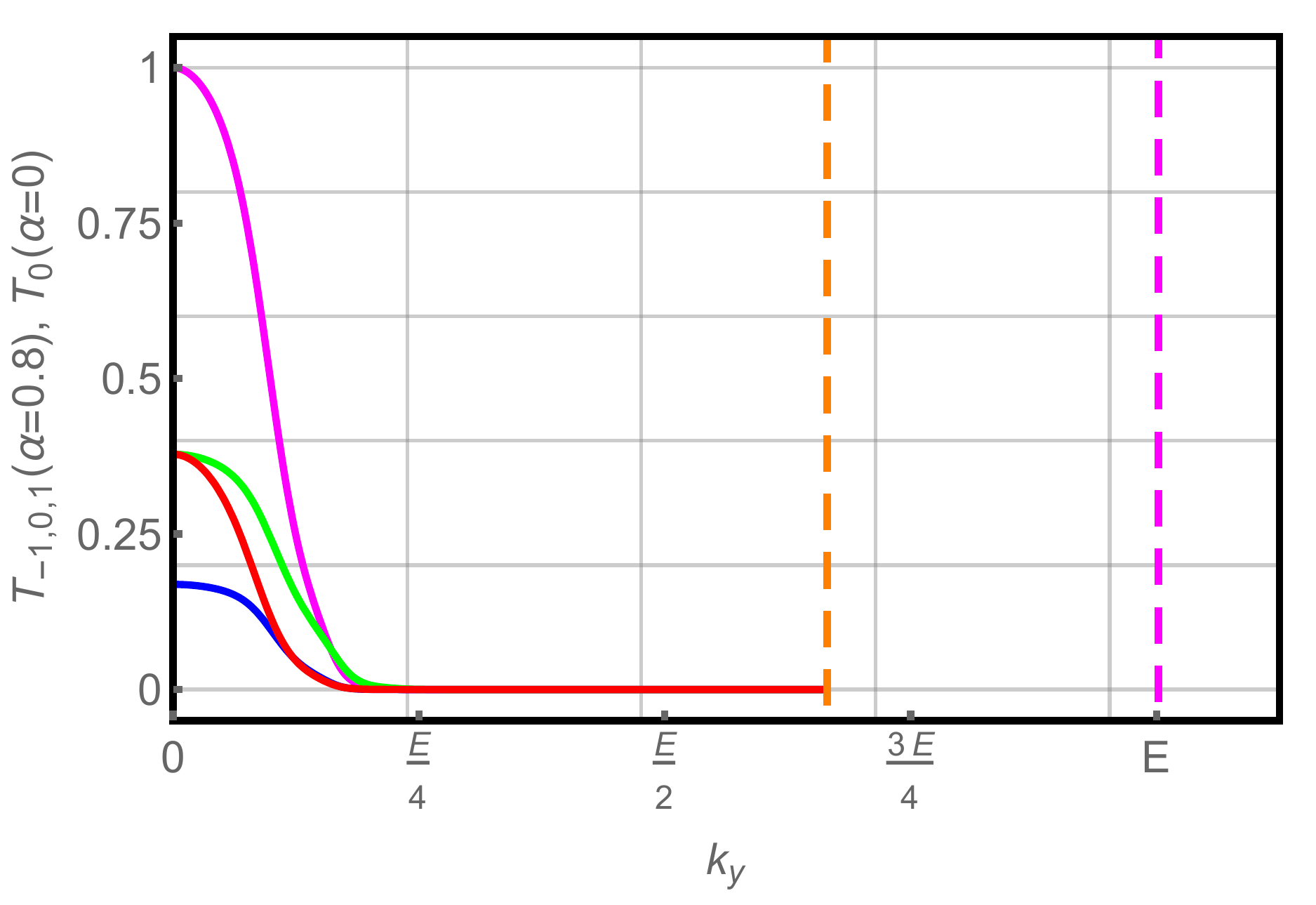}
    \label{FigCh04:403}
}\\
\subfloat[]{
    \includegraphics[scale=0.28]{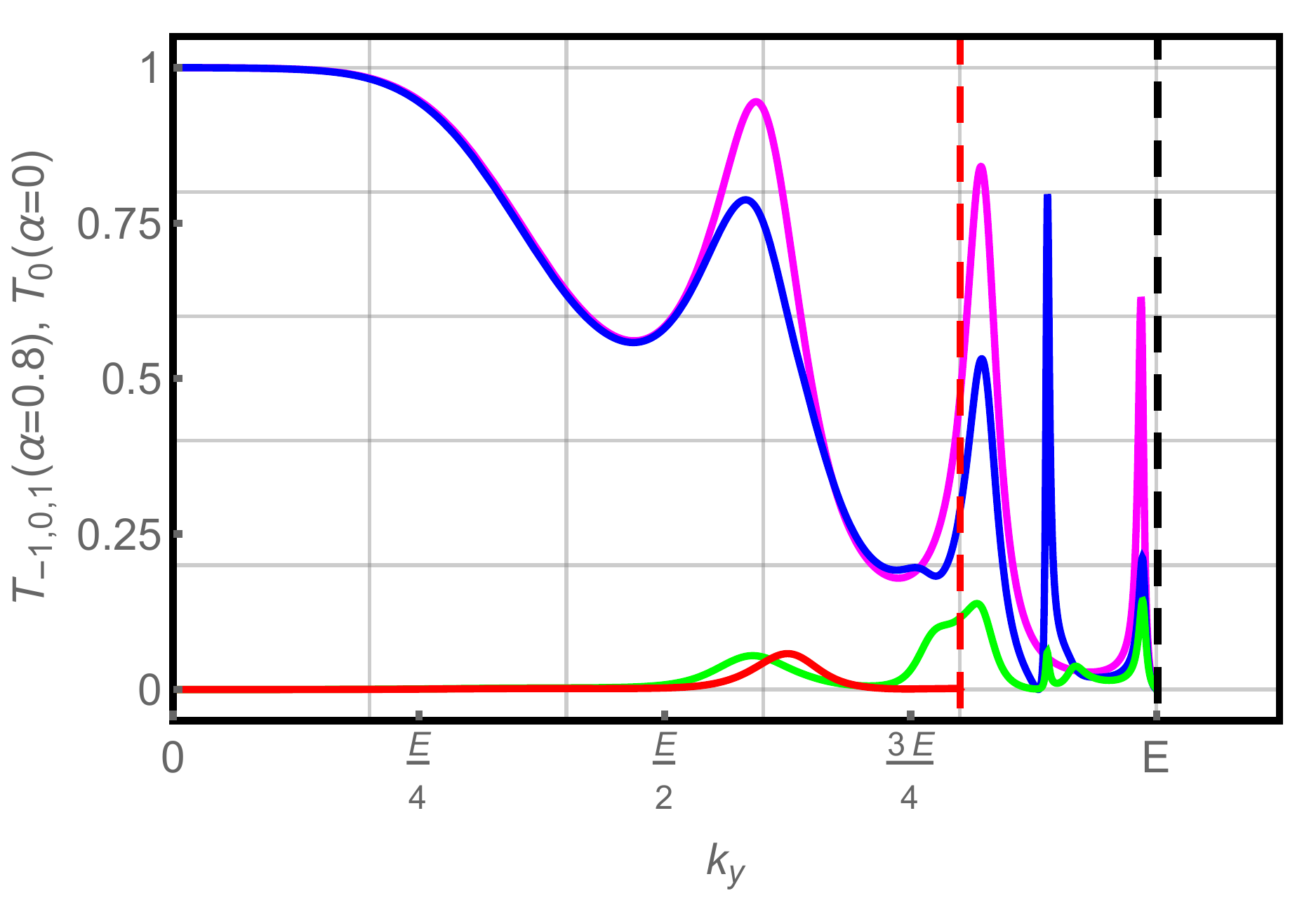}
    \label{FigCh04:404}
}
\subfloat[]{
    \includegraphics[scale=0.28]{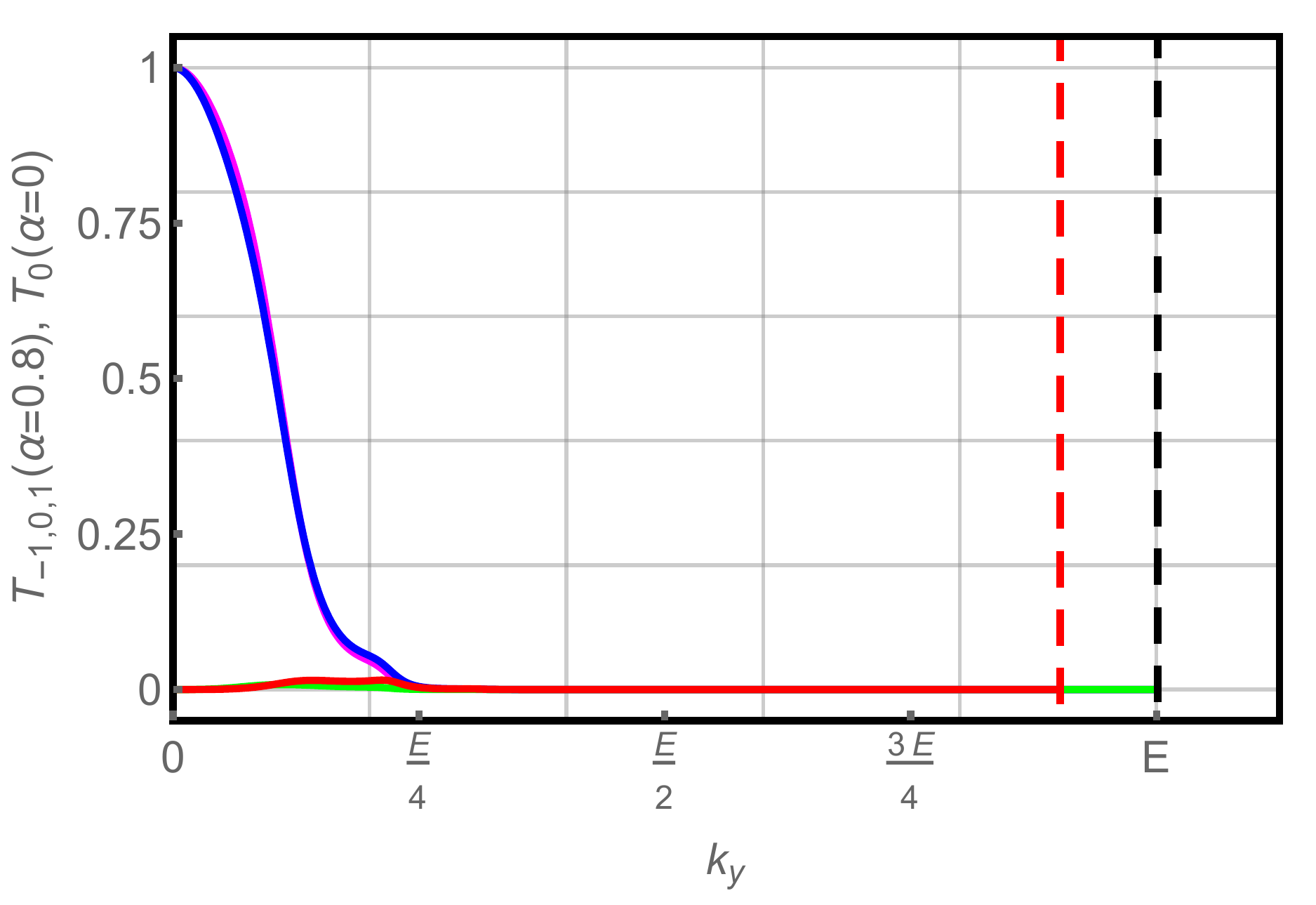}
    \label{FigCh04:405}
}
\subfloat[]{
    \includegraphics[scale=0.28]{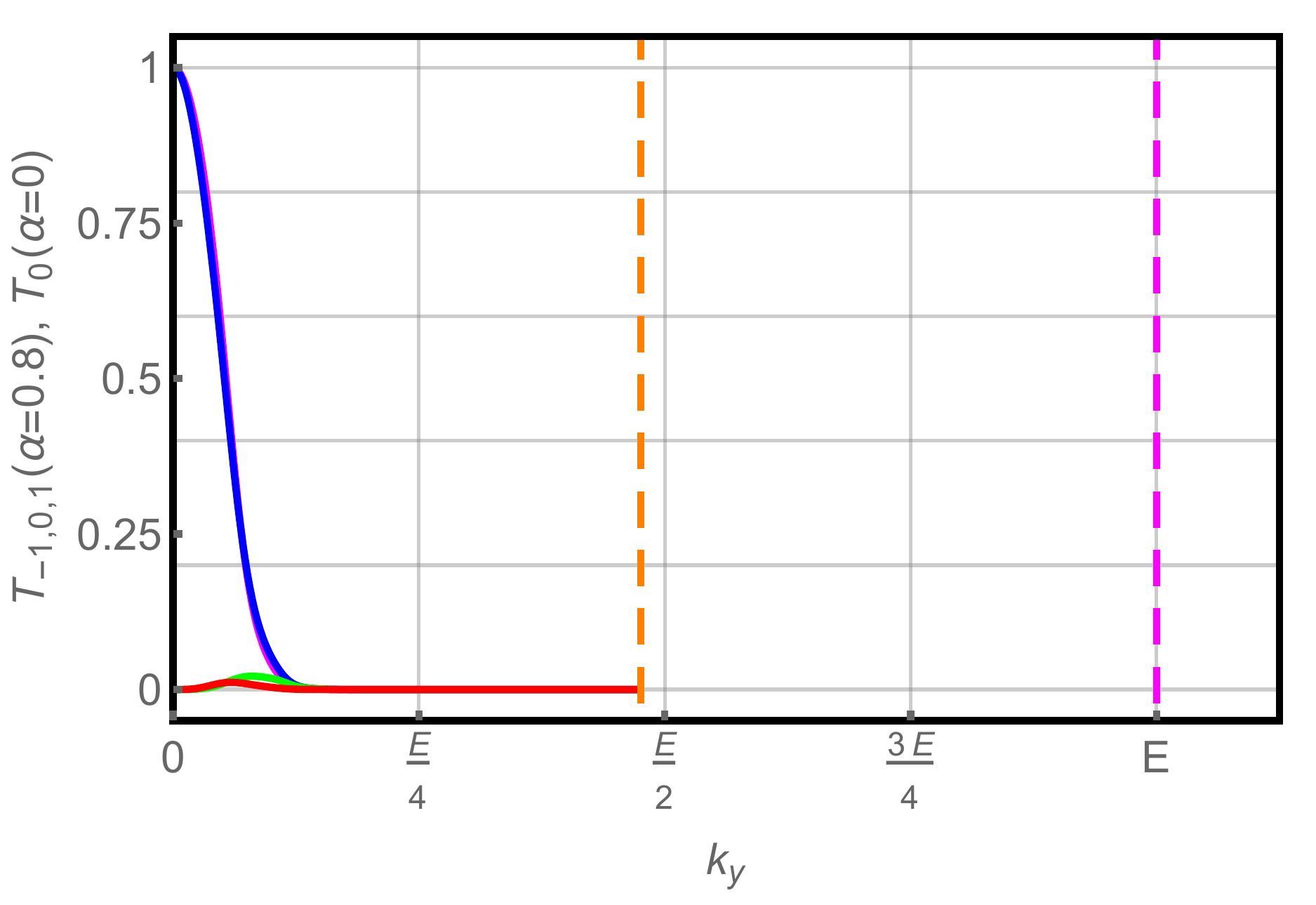}
    \label{FigCh04:406}
}
\caption{(Color online) Transmission probabilities versus  wave vector $k_y$ 
with  $\alpha=0,0.8$, $V_1=10$, $V_0=20$, $T_{0}(\alpha=0)$ 
(magenta color), $T_{0}(\alpha=0.5)$ (blew color), $T_{-1}(\alpha=0.5)$ (red color), and $T_{1}(\alpha=0.5)$ 
(green color),  for  \protect\subref{FigCh04:401}: $E=5$ and $L=\pi$, 
\protect\subref{FigCh04:402}: $E=10$ and $L=\pi $, \protect\subref{FigCh04:403}: $E=20$ and $L=\pi$, 
\protect\subref{FigCh04:404}: $E=5$ and $L=2\pi $, \protect\subref{FigCh04:405}: $E=10$ and $L=2\pi$, 
\protect\subref{FigCh04:406}: $ E=20$ and $L=2\pi$.}
\label{FigCh44}
\end{figure}

{Figure \ref{FigCh44} shows the transmission probabilities as function of wave vector $k_y$
under suitable conditions. To describe such Figure
we divide the interval of $k_y$
into two 
regions:  
 $\Big\{R_1=]0,k_{y_1}[$, $R_2=]k_{y_1},k_{y_2}[$ 
 (with $k_{y_1}$: dashed red line, $k_{y_2}=E$: dashed black line)$\Big\}$
corresponding to Figures (\ref{FigCh44}\subref{FigCh04:401}, \ref{FigCh44}\subref{FigCh04:402},
\ref{FigCh44}\subref{FigCh04:404},
\ref{FigCh44}\subref{FigCh04:405})
and 
 $\Big\{R_3=]0,k_{y_3}[$, 
 $R_4=]k_{y_3},k_{y_4}[$  (with $k_{y_3}<E$: dashed orange line, 
$k_{y_4}=E$: dashed magenta line)$\Big\}$ corresponding to Figures 
(\ref{FigCh44}\subref{FigCh04:403}, \ref{FigCh44}\subref{FigCh04:406}). 
We observe that for any value of energy,   always there is the condition 
 $ k_y <E $.
 In} $R_1$ and $R_3$ there is manifestation of all transmission  modes, but 
 in $R_2$ the $T_{-1}(\alpha=0.8)$ mode is missing, and in $R_4$ all modes 
 $(T_{-1,0,1}(\alpha=0.8),T_{0}(\alpha=0))$ are  forbidden.
In $R_1$ (Figure \ref{FigCh44}\subref{FigCh04:401}), $T_{0}(\alpha=0)$ is the first dominant, 
followed by $T_{-1,1}(\alpha=0.8)$, which are co-dominant at the beginning 
and finish by dominance of $T_{1}(\alpha=0.8)$ with respect to $T_{-1}(\alpha=0.8)$.
Finally the central
band $T_{0}(\alpha=0)$ dominates $T_{-1}(\alpha=0.8)$ only just 
before $k_{y_1}$. In $R_2$ (Figure \ref{FigCh44}\subref{FigCh04:401}), 
$T_{0}(\alpha=0)$, $T_{1}(\alpha=0.8)$, $T_{0}(\alpha=0.8)$ show peaks 
(one peak, two peaks, two peaks) respectively, the first peak of $T_{0}(\alpha=0)$ dominates 
the first peak of $T_{0}(\alpha=0.8)$ that also dominates the first  peak of $T_{1}(\alpha=0.8)$
and 
the second peak of $T_{0}(\alpha=0.8)$ dominates the second peak of $T_{1}(\alpha=0.8)$. At the 
beginning of  $R_1$ (Figure \ref{FigCh44}\subref{FigCh04:404}), 
($T_{0}(\alpha=0)$, $T_{0}(\alpha=0.8)$ are co-dominant between them and dominate together 
the two transmissions $T_{-1,1}(\alpha=0.8)$ which are also co-dominant between them. After 
$k_y=\frac{E}{2}$, each of the four transmissions generates a first peak, where that 
of $T_{0}(\alpha=0)$  is greater than that of $T_{0}(\alpha=0.8)$,  which also greater than that of 
$T_{-1,1}(\alpha=0.8)$. The transmissions in  $R_2$ (Figure \ref{FigCh44}\subref{FigCh04:404}) behave
in the same way as those in  $R_2$ (Figure \ref{FigCh44}\subref{FigCh04:401}), except towards the 
end of region one has a generation of a supplementary peak for each transmission with 
$T_{0}(\alpha=0)>T_{0}(\alpha=0.8)>T_{1}(\alpha=0.8)$. 
{In  
$R_1$ (Figure \ref{FigCh44}\subref{FigCh04:402}), $R_3$ (Figure \ref{FigCh44}\subref{FigCh04:403}), 
$R_1$ (Figure \ref{FigCh44}\subref{FigCh04:405}), and $R_3$ (Figure \ref{FigCh44}\subref{FigCh04:406}), 
each transmission probability has one peak}. We observe that there are {orders:
 $T_{0}(\alpha=0)>T_{-1}(\alpha=0.8)>T_{1}(\alpha=0.8)> T_{0}(\alpha=0.8)$ in
$R_1$ (Figure \ref{FigCh44}\subref{FigCh04:402}), 
$T_{0}(\alpha=0)>T_{1}(\alpha=0.8)>T_{-1}(\alpha=0.8)> T_{0}(\alpha=0.8)$ in  $R_3$ 
(Figure \ref{FigCh44}\subref{FigCh04:403}),
$T_{0}(\alpha=0)=T_{0}(\alpha=0.8)>T_{0-1}(\alpha=0.8)\approx T_{1}(\alpha=0.8)$
in  $R_1$ (Figure \ref{FigCh44}\subref{FigCh04:405}) 
and $R_3$ (Figure \ref{FigCh44}\subref{FigCh04:406}).}


\section{Conclusion}

{We have considered a system composed of three regions of 
graphene  where the intermediate one was subjected to 
a linear potential barrier oscillating  in time
 with driving frequency $\omega$ along $x$-direction. The infinite mass boundary condition were used
to quantize the wave vector $k_y$ along $y$-direction. Due to separability of the
eigenspinors we have applied the Floquet theorem to obtain the energy side-bands.
By solving the eigenvalue equation,  the solutions of the energy spectrum 
for each region were derived.
Moreover,
it is showed
that
the  barrier in time generated an infinite modes giving rise to
several energy modes.
}

{Subsequently,
 the transport properties of 
the present system through transmission probabilities was studied 
using the transfer matrix approach. Indeed, after matching the eigenspinors at interfaces
we have calculated the corresponding transmission and reflections coefficients. 
There were used together with the current density to explicitly determine the transmission and 
reflection probabilities as function of the physical parameters. For numerical limitation, we have showed
%
that}
how the transmission probabilities for the central
band ($l=0$) and two first sidebands ($l=\pm 1$) are affected by various physical
parameters such that  incident energy $E$,  barrier width $L$
and transverse wave vector $k_y$. {These were supported by offering
different plots exhibiting the transmission behavior for three side-bands $l=0,\pm 1$}.

In summary, our numerical results support the
assertion that quantum interference has an important effect on
fermions through graphene based linear
 barrier driven by  time-oscillations. Since most optical applications in electronic devices are based
on interference phenomena then we expect that the results of our
computations might be of interest to designers of graphene-based
electronic devices.

\section*{Acknowledgments}
The generous support provided by the Saudi Center for Theoretical Physics (SCTP)
is highly appreciated by all authors. 

\section*{Author contribution statement}
All authors contributed equally to the paper.

\end{document}